\pdfoutput=1

\documentclass[times]{speauth}
\usepackage{colortbl}
\usepackage{listings}
\usepackage{verbatim}
\usepackage{tikz}
\usepackage{color}

\usetikzlibrary{chains,positioning,arrows}

\usepackage{ifthen}
\newcommand{\deanonymizedversion}{show them}       % comment this line to deanonymize

\ifthenelse{\isundefined{\deanonymizedversion}}{%
 \newenvironment{anonymous}{\expandafter\comment}{\expandafter\endcomment}

\newcommand{\anonymizeurl}[1]{\url{anonymized}}
\pdfinfo{%
  /Title (SIMD Compression and the Intersection of Sorted Integers)
  /Creator ()
  /Producer ()
  /Author ()
  /Subject ()
  /Keywords ()
}
}{%
 \newenvironment{anonymous}{}{}

\newcommand{\anonymizeurl}[1]{\url{#1}}
\pdfinfo{%
  /Title (SIMD Compression and the Intersection of Sorted Integers)
  /Creator (Daniel Lemire, Leonid Boytsov, Nathan Kurz)
  /Producer ()
  /Author ()
  /Subject ()
  /Keywords ()
}

}

\newcommand{\Tau}{\mathrm{T}}
\usepackage{todonotes}
\usepackage{paralist}

\usepackage[colorlinks,bookmarksopen,bookmarksnumbered,citecolor=red,urlcolor=red]{hyperref}
 \usepackage{siunitx}
\usepackage{subfloat,subfig,float,graphicx,verbatim,url}
\usepackage{multicol,multirow}
\usepackage{arydshln}
\usepackage{algorithm}
\usepackage[noend]{algorithmic} % can use noend to play with spacing

\usepackage{pifont}

\usepackage{flushend}

\usepackage{todonotes}

\DeclareSIUnit \GHz {\giga{}\hertz{}}
\DeclareSIUnit \byte {B}
\DeclareSIUnit \MB {MB}
\DeclareSIUnit \kB {kB}
\DeclareSIUnit \GB {GB}

\newcommand*\BitAnd{\mathrel{\&}}
\newcommand*\BitOr{\mathrel{|}}
\newcommand*\ShiftLeft{\ll}
\newcommand*\ShiftRight{\gg}

\usepackage{enumitem}

\graphicspath{{./figures/}}

\begin{document}

\runningheads{D.~Lemire, L. Boytsov, N.~Kurz}{SIMD Compression and the Intersection of Sorted Integers}

\title{SIMD Compression and the Intersection\\ of Sorted Integers}

\author{D.~Lemire\affil{1}\corrauth,  L. Boytsov\affil{2}, N. Kurz\affil{3}}

\address{\affilnum{1}LICEF Research Center, TELUQ, Montreal, QC, Canada\break
\affilnum{2}Carnegie Mellon University, Pittsburgh, PA USA\break
\affilnum{3}Verse Communications, Orinda, CA USA %
}

\corraddr{LICEF Research Center, TELUQ, Universit\'e du Qu\'ebec, 5800 Saint-Denis,  Montreal (Quebec) H2S 3L5 Canada.}
\cgsn{Natural Sciences and Engineering Research Council of Canada}{261437}

%
%\numberofauthors{3} %  in this sample file, there are a *total*
%\ifthenelse{\isundefined{\deanonymizedversion}}{%
%\author{%
%\alignauthor%
%~\\
%       \affaddr{~}\leavevmode\\
%       \affaddr{~}\leavevmode\\
%       \affaddr{~}\leavevmode\\
%       \affaddr{~}\leavevmode\\
%       \email{~}
%}
%}{%
%\author{%
%\alignauthor
%{Daniel Lemire}\leavevmode\\
%       \affaddr{{LICEF Research Center}}\leavevmode\\
%       \affaddr{{TELUQ, Universit\'e du Qu\'ebec}}\leavevmode\\
%       \affaddr{{5800 Saint-Denis}}\leavevmode\\
%       \affaddr{{Montreal, QC H2S 3L5 Can.}}\leavevmode\\
%       \email{{lemire@gmail.com}}
%\alignauthor
%{Leonid Boytsov}\leavevmode\\
%       \affaddr{\mbox{{Language Technologies Institute}}}\leavevmode\\
%       \affaddr{{Carnegie Mellon University}}\leavevmode\\
%       \affaddr{{5000 Forbes Ave}}\leavevmode\\
%       \affaddr{{Pittsburgh, PA 15213, USA}}\leavevmode\\
%       \email{{srchvrs@cmu.edu}}
%%% 2nd. author
%\alignauthor
%{Nathan Kurz}\leavevmode\\
%       \affaddr{{Verse Communications}}\leavevmode\\
%       \affaddr{{84 La Espiral}}\leavevmode\\
%       \affaddr{{Orinda, CA 94563 USA}}\leavevmode\\
%       \email{{nate@verse.com}}
%%% 3rd. author
%}
%}
%
%\maketitle
\begin{abstract}
Sorted lists of integers are commonly used in inverted indexes and database systems. They are often compressed in memory.
We can use the SIMD instructions available in common processors to boost the speed of integer compression schemes. Our S4-BP128-D4 scheme uses as little as 0.7~CPU cycles per decoded 32-bit integer
while still providing state-of-the-art compression.

However, if the subsequent processing of the integers is slow, the effort spent on optimizing decompression speed can be wasted. To show that it does not have to be so, we \begin{inparaenum}[(1)]\item vectorize and optimize the intersection of posting lists; \item introduce the \textsc{SIMD Galloping} algorithm. \end{inparaenum}
We exploit the fact that one SIMD instruction can compare 4~pairs of 32-bit integers at once.

% no more AOL
We experiment with two TREC text collections, GOV2 and ClueWeb09 (Category B), using
logs from the
TREC million-query track. We show that  using only the SIMD instructions ubiquitous in all modern CPUs,  our techniques for conjunctive queries can double the speed of a state-of-the-art approach.

\end{abstract}

%\category{E.4}{Coding and Information Theory}{Data Compaction and Compression}
%\category{D.2.8}{Software Engineering}{Metrics}[complexity measures, performance measures]


%\keywords{performance; measurement; index compression; vector processing}
\keywords{performance; measurement; index compression; vector processing}

\maketitle
%\todo{
%D4. The terms ``unpacking'' and ``decoding'' do not carry consistent meanings throughout the text. Sometimes they mean decompressing the bit-packed formats in Section 3, sometimes they mean computing the prefix-sum in Section 4, and sometimes they mean the two combined, which causes much confusion! For example, the ``unpacking'' of Section 6.4 and ``decoding'' of Section 6.5 seem to refer to the same thing.}

\section{Introduction}

An inverted index maps terms to lists of document identifiers.
A column index in a database might, similarly, map attribute values to row identifiers. Storing all these lists on disk can limit the performance since the latency of the fastest drives is several orders of magnitude higher than the latency of memory.
% Thus, engineers commonly compress these lists of integers so that they fit in memory.
Fast compression can reduce query response times~\cite{Catena2014,Buttcher:2007:ICG:1321440.1321546}.

We assume that identifiers can be represented using 32-bit integers and that they are stored in sorted order. In this context,
one can achieve good compression ratios and high decompression speed, by employing \emph{differential coding}.
We exploit the fact that the lists are sorted and instead of storing the integers themselves,
we store the differences between successive integers, sometimes called the deltas.

Additional improvements arise from using the single instruction,
multiple data (SIMD) instructions available on practically all server and desktop processors produced in
the last decade (Streaming SIMD Extensions~2 or SSE2).
A SIMD  instruction performs the same operation on
multiple pieces of data: this is also known as vector processing.
Previous research~\cite{LemireBoytsov2013decoding} showed that we  can decode compressed 32-bit integers using less than 1.5~CPU cycles per integer in a realistic inverted index scenario by using  SIMD instructions. We expect that this is at least twice as fast as any non-SIMD scheme.

One downside of differential coding is that decompression requires the computation of a \emph{prefix sum} to recover the original integers~\cite{Ladner:1980:PPC:322217.322232}: given the delta values $\delta_2, \delta_3, \ldots$ and an initial value $x_1$, we must compute $x_1+\delta_2, x_1+\delta_2+\delta_3, x_1+\delta_2+\delta_3+\delta_4+\ldots$.
When using earlier non-SIMD compression techniques, the computational cost of the prefix sum might be relatively small,
but when using faster SIMD compression, the prefix sum can account for up to half of the running time. Thankfully we can accelerate the computation of the prefix sum using SIMD instructions.

Our first contribution is to revisit the computation of the prefix sum. On 128-bit SIMD vectors, we introduce four~variations exhibiting various speed/compression trade-offs, from the most common one, to the 4-wise approach proposed  by Lemire and Boytsov~\cite{LemireBoytsov2013decoding}. In particular, we show that at the same compression ratios, vector
(SIMD) decompression is much faster  than scalar (non-SIMD)
decompression.
Maybe more importantly, we show that \emph{integrating} the prefix sum with the unpacking can nearly double the speed. Some of our improved
schemes can decompress more than one 32-bit integer per CPU cycle. We are not aware of any similar high speed previously reported, even accounting for hardware differences.

To illustrate that SIMD instructions can significantly benefit other aspects of an inverted index (or a database system), we consider the problem of computing conjunctive queries: e.g., finding all documents that contain a given set of terms.
In some search engines, such conjunctive queries are the first, and, sometimes, the most expensive, step in query processing~\cite{Culpepper:2010:ESI:1877766.1877767,Tatikonda:2011:PLI:2009916.2010045}.
 Some categories of users, e.g., patent lawyers~\cite{Kim2011}, prefer to use complex Boolean queries---where conjunction is an important part of query processing.

 Culpepper and Moffat showed that a competitive approach (henceforth \textsc{hyb+m2}) was to represent the longest lists as
 bitmaps while continuing to compress the short lists
using differential coding~\cite{Culpepper:2010:ESI:1877766.1877767}.
  To compute an intersection, the short lists are first intersected and then the corresponding bits in the bitmaps are read to finish the computation.
As is commonly done, the short lists are intersected two-by-two  starting from the two smallest lists: this is often called a Set-vs-Set or SvS processing.
These intersections are typically computed using scalar algorithms.
 In fact, we are not aware of any SIMD intersection algorithm proposed for such problems on CPUs. Thus as our second contribution, we introduce new SIMD-based intersection algorithms for uncompressed integers that are up to twice as fast as competitive scalar intersection algorithms. By combining both the fast SIMD decompression  and fast SIMD intersection,
 we can double the speed of the intersections
on compressed lists.

We summarize our main contributions as follows:
\begin{enumerate}
\item We show that by combining SIMD unpacking and the prefix sum required by   differential coding, we can improve decompression speed by up to \SI{90}{\percent}. We end up beating by about \SI{30}{\percent} the previous best reported decompression speeds~\cite{LemireBoytsov2013decoding}.
\item We introduce a  family of  intersection algorithms to exploit commonly available SIMD instructions (\textsc{V1}, \textsc{V3} and \textsc{SIMD Galloping}). They are often twice as fast as the best non-SIMD algorithms.
\item By combining our results, we  nearly double the speed of a fast  inverted index (\textsc{hyb+m2}) over realistic data and queries---on standard PC processors.
\end{enumerate}

To ease reproducibility, we make all of the software and data sets publicly available,
including the software to process the text collections and generate query mappings (see \S~\ref{SectSoftware} and \S~\ref{SectRealData}).

\section{Related Work}\label{sec:relatedwork}

For an exhaustive review of fast 32-bit integer compression techniques, we refer the reader to Lemire and Boytsov~\cite{LemireBoytsov2013decoding}. Their main finding is that schemes compressing integers in large ($\approx 128$) blocks of integers with minimal branching are faster than other approaches, especially when using SIMD instructions. They reported using fewer than 1.5~CPU cycles per 32-bit integer
on a 2011-era Intel \emph{Sandy Bridge} processor. In comparison, Stepanov et al.~\cite{Stepanov:2011:SDP:2063576.2063627} also proposed compression schemes optimized for SIMD instructions on CPUs, but they reported using at least 2.2~CPU cycles per 32-bit integer on a 2010 Intel \emph{Westmere} processor.

Regarding the intersection of sorted integer lists, Ding and K\"onig~\cite{Ding:2011:FSI:1938545.1938550} compared a wide range of algorithms
in the context of a search engine and found that SvS with \emph{galloping} was competitive: we describe galloping in \S~\ref{sec:fastintersection}. Their own technique (RanGroup) performed better ($\approx$\SI{10}{\percent}--\SI{30}{\percent}) but it does not operate over sorted lists but rather over a specialized data structure that divides up
the data randomly into small chunks.
In some instances, they found that a merge (akin to the merge step in the merge sort algorithm) was faster. They also achieved  good results with Lookup~\cite{Transier:2010:EBA:1877766.1877768}: a technique that relies on an auxiliary data structure for skipping values~\cite{moffat1996self}.
%\leo{Is ``skipping ahead'' a correct term? I am not quite sure.}
%\daniel{Rephrased as skipping values which is the exact terminology used by Transier.}
%Leo ok
Ding and K\"onig found that  alternatives such as
Baeza-Yates' algorithm~\cite{springerlink:10.1007/978-3-540-27801-6_30}
or adaptive algorithms~\cite{Demaine:2000:ASI:338219.338634} were slower.
%\leo{Is Baeza-Yates' algorithm adaptive?}

%\daniel{I do not think that Baeza-Yates has been described as being adaptative. I do not think that Baeza-Yates presents it this way.}
%Leo ok

Barbay et al.~\cite{Barbay:2010:EIS:1498698.1564507} also carried out an extensive experimental evaluation. On synthetic data using a uniform distribution, they found that Baeza-Yates' algorithm~\cite{springerlink:10.1007/978-3-540-27801-6_30} was faster than SvS with galloping (by about \SI{30}{\percent}). However, on real data (e.g., TREC GOV2), SvS with galloping was superior to most alternatives by a wide margin (e.g., $2 \times$ faster).

Culpepper and Moffat similarly found that SvS with galloping was the fastest~\cite{Culpepper:2010:ESI:1877766.1877767} though their own \texttt{max} algorithm was fast as well. They found that in some specific instances (for queries containing 9 or more terms) a technique similar to galloping (interpolative search) was slightly better (by less than \SI{10}{\percent}).

Kane and Tompa improved Culpepper and Moffat's \textsc{hyb+m2} by adding auxiliary data structures to skip over large blocks of compressed values (256~integers) during the computation of the intersection~\cite{kane2014}. Their good results are in contrast with  Culpepper and Moffat's finding that skipping is counterproductive when using bitmaps~\cite{Culpepper:2010:ESI:1877766.1877767}.
%Their approach could be combined with our own ideas, in future research.
%\leoinline{Whose approach can be combined with whose ideas? BTW, it seems you already tried to apply their good ideas (in the appendix). This showed ideas to be ineffective. Did I miss something?}
%%% Daniel: I think that the text was clear that we refereed to Kane and Tompa's results but in light of our new appendix, let us kill this last bit, thus solving the problem.
%% ok, it was confusing at a higher level. We agree that skipping can be productive, but we fail to show this. In addition, we reference other people who showed the same thing.

Our work is focused on commodity desktop processors. Compression and intersection of integer lists using a graphics processing unit (GPU) has also received attention.
Ding et al.~\cite{Ding:2009:UGP:1526709.1526766} improved the intersection speed using a \emph{parallel merge find}: essentially, they divide up one list into small blocks and intersect these blocks in parallel with the other array.
On conjunctive queries, Ding et al.~\cite{Ding:2009:UGP:1526709.1526766} found their GPU implementation to be only marginally superior to a CPU implementation ($\approx$\SI{15}{\percent} faster) despite the data was already loaded in GPU's global memory. They do, however, get impressive speed gains ($7 \times$) on disjunctive queries.

Ao et al.~\cite{Ao:2011:EPL:2002974.2002975} proposed a parallelized compression technique (Parallel PFor) and replaced conventional differential coding with an approach based on linear regression.
In our work, we rely critically on differential coding, but  alternative models merit
consideration~\cite{Konow2013,Vigna2013,Ottaviano2014}.

\section{Relevant SIMD instructions}\label{sec:simdinstructions}

Intel PC processors support vector instructions, but languages such as C or C++ do not directly include vectorization as part of their standard syntax. However, it is still possible to conveniently call these instructions by using intrinsics  or online assembly code.
Intrinsics are special functions (sometimes called built-in functions) provided as a compiler extension of C/C++ syntax.

Table~\ref{ref:simdinstructions} presents the various SIMD instructions we require. In addition
to instruction mnemonic names, we also provide names of respective C/C++ intrinsics.
In this table, there is a single intrinsic function per mnemonic except for the move instruction
\texttt{movdqu}, which has separate intrinsics to denote store and load operations.

Though most of the instructions are fast, some are significantly more expensive. Intel often expresses the computational cost of an instruction in terms of its latency and reciprocal throughput. The latency is the minimum number of cycles required to execute the instruction. The reciprocal throughput is one over  the maximum number of instructions of the same kind that can be executed per cycle. For example, a reciprocal throughput of 0.5 indicates that up to two instructions of the same type can be executed in a cycle. We give the latency and reciprocal throughput
for Intel processors with a \emph{Sandy Bridge} microarchitecture~\cite{fog2014instruction}. These numbers are often  constant across Intel microarchitectures.

All of these instructions use 128-bit registers (called \emph{XMM registers}). Many of them are straightforward. For example, \texttt{por} and \texttt{pand} compute the bitwise OR and AND between two registers.

We use the \texttt{movdqu} instruction to load or store a register. Loading and storing registers has a relatively high latency (3~cycles). While we can load two registers per cycle, we can only store one of them to memory.

The bit shifting instructions come in two flavors. When we need to shift entire registers by a number of bits divisible by eight (a byte), we use the \texttt{psrldq} and \texttt{pslldq} instructions. They have a high throughput (2~instructions per cycle) on \emph{Sandy Bridge} Intel processors~\cite{intelintrin}.
%\leoinline{

%Interestingly, the manual that you reference doesn't seem to support this claim.

%Maybe, I am missing something, but check pages 166 and 196.

%I also found utility to measure \url{https://github.com/aizvorski/mubench}.

%On fastpfor, it shows that pslldq, pshufd, and pshufb have the reciprocal throughput of 0.5,

%but on my core i7 laptop the reciprocal throughput of all three operations seems to be 1.

%The throughput of \texttt{psrld} is measured to be 1 on both computers.

%Surprisingly, despite these differences,

%there is no visible difference in performance between my laptop and fastpfor for all applications of interest.

%If you ever use mubench, please, specify the CPU frequency explicitly.

%It seems to fail to detect it properly otherwise.}
%\danielinline{If I recall, your laptop is haswell, not sandy bridge. I have made the statement more precise here by saying that we only claim this about Sandy Bridge. As your tests suggest, it is an accurate statement for Sandy Bridge. I have also added a reference to the Intel documentation where this information is provided. }
%great reference (all the latencies are easy to find)!!!
 We can also consider the 128-bit registers as a vector of four 32-bit integers. We can then shift right four 32-bit integers by a fixed number of bits using the \texttt{psrld} instruction. It has reduced throughput compared to the byte-shifting instructions (1~instruction per cycle).

In the same spirit, we can add four 32-bit integers with four other 32-bit integers at once using the \texttt{paddd} instruction. It is another fast operation.

Sometimes it is necessary to copy the 32-bit integers from one XMM register to another, while possibly moving or duplicating values. The \texttt{pshufd} instruction can serve this purpose.
It takes as a parameter an input register $v$ as well as a control mask $m$. The control mask is made of four 2-bit integers each representing an integer in $\{0,1,2,3\}$. We output $(v_{m_0},v_{m_1},v_{m_2},v_{m_3})$.
%with the convention that $v_{-1}\equiv 0$. \leo{What is -1 here?}
%% Daniel: Unecessary garbage having to do with pshufb
%%ok
Thus, for example, the  \texttt{pshufd} instruction can copy one particular value to all positions (using a mask made of 4~identical values). It is a fast instruction with a throughput of two instructions per cycle.

We can compare the four 32-bit integers of one register $v$ with the four 32-bit integers of another register $v'$ using the \texttt{pcmpeqd} instruction. It  generates four 32-bit integers with value 0xFFFFFFFF or 0 depending on whether the corresponding pairs of integers are equal (e.g., if $v_0=v'_0$, then the first component returned by \texttt{pcmpeqd} has value 0xFFFFFFFF).

The \texttt{pcmpeqd} instruction can be used in tandem with the \texttt{movmskps} instruction,
 which generates a 4-bit mask by extracting the four most significant bits from the four 32-bit integers produced by the comparison instruction \texttt{pcmpeqd}. The \texttt{movmskps} instruction is slightly  expensive, with a throughput of 1~instruction per cycle and a latency of 2~cycles.
In some cases, we only need to verify if four 32-bit integers in one register
are different from respective 32-bit integers in another register, i.e., we do not need to extract
a mask.
To check this, we use either the SSE2 instruction \texttt{pmovmskb} or
the SSE4 instruction \texttt{ptest}.
The SSE4 instruction has smaller latency, however, replacing
\texttt{ptest} with \texttt{pmovmskb} did not substantially affect runtime.

%\leoinline{I replaced SSE4.1 with SSE4, because some of the instructions are actually SSE4.2}
%% Daniel: Ok. Good.
In some particular algorithms (see \S~\ref{sec:existing}), we also use two recently introduced string-comparison instructions (part of the SSE4 instruction set):
\texttt{pcmpestrm}  and \texttt{pcmpistrm}. They operate on 8-bit or 16-bit strings loaded in XMM registers. They take a control mask to specify their behavior.
We use them for 16-bit strings.
 For our purposes, we ask for a 8-bit mask
indicating whether any of the eight 16-bit elements of the first register are equal to any of the eight 16-bit elements of the second register.
This mask is stored in a 128-bit MMX register and is extracted using the \texttt{pextrd} instruction.

The two string-comparison instructions differ in how
they deduce the string length. The slower
\texttt{pcmpestrm} instruction requires us to specify the string lengths (8~elements in our case). The faster  \texttt{pcmpistrm} instruction assumes that strings are null terminated. When no null value is found, the \texttt{pcmpistrm} instruction processes 8~elements per register.
There is no 16-bit counterpart to the
\texttt{pshufd} instruction, but there is an 8-bit version (\texttt{pshufb}) with the same latency and throughput.

\begin{table}
\caption{SIMD instructions on  \emph{Sandy Bridge} Intel processors with latencies and reciprocal throughput in CPU cycles \label{ref:simdinstructions}.
All instructions are part of SSE2, except
\texttt{pcmpestrm}, \texttt{pcmpistrm}, \texttt{pextrd}, \texttt{popcnt}, \texttt{ptest} that are SSE4 instructions,
and \texttt{pshufb} that is an SSSE3 instruction.}\centering
\begin{tabular}{ccp{2.0in}cc}
\toprule
instruction & C/C++ intrinsic & description & latency & rec.\ thr.
\\\midrule
\small\texttt{por}  & \small\texttt{\_mm\_or\_si128} & bitwise OR & 1 & 0.33\\
\small\texttt{pand} & \small\texttt{\_mm\_and\_si128} & bitwise AND & 1 & 0.33\\
\small\texttt{movdqu} & \small\texttt{\_mm\_storeu\_si128} & store a 128-bit register & 3 & 1 \\
\small\texttt{movdqu} & \small\texttt{\_mm\_loadu\_si128} & load to 128-bit register & 3 & 0.5\\
\small\texttt{psrldq} & \small\texttt{\_mm\_srli\_si128} & shift right by a number of bytes & 1 & 0.5\\
\small\texttt{pslldq}  &\small\texttt{\_mm\_slli\_si128} & shift left by a number of bytes & 1 & 0.5 \\
\small\texttt{psrld}  &\small\texttt{\_mm\_srl\_epi32} & shift right four 32-bit integers & 1 & 1\\
\small\texttt{pslld}  &\small\texttt{\_mm\_sll\_epi32} & shift left four 32-bit integers & 1 & 1\\
\small\texttt{paddd}  &\small\texttt{\_mm\_add\_epi32} & add four 32-bit integers & 1 & 0.5\\
\small\texttt{pshufd}  &\small\texttt{\_mm\_shuffle\_epi32} & shuffle four 32-bit integers & 1 & 0.5\\
\small\texttt{pcmpeqd}  &\small\texttt{\_mm\_cmpeq\_epi32} & compare four 32-bit integers for equality & 1 & 0.5\\
\small\texttt{movmskps}  &\small\texttt{\_mm\_movemask\_ps} & mask from most significant bits of 32-bit elements & 2 & 1\\
\small\texttt{pmovmskb}  &\small\texttt{\_mm\_movemask\_epi8} & mask from most significant bits of 16-bit elements & 2 & 1\\
\midrule
\small\texttt{pcmpestrm}  &\small\texttt{\_mm\_cmpestrm} & compare two strings of specific lengths & 12 & 4\\
\small\texttt{pcmpistrm}  &\small\texttt{\_mm\_cmpistrm} & compare two null-terminated strings & 11 & 3\\
\small\texttt{pshufb}  &\small\texttt{\_mm\_shuffle\_epi8} & shuffle 16 bytes & 1 & 0.5\\
\small\texttt{pextrd}  &\small\texttt{\_mm\_extract\_epi32} & extract a specified 32~bit integer & 2 & 1 \\
\small\texttt{popcnt}  &\small\texttt{\_mm\_popcnt\_u32} & number of 1s in a 32-bit integer & 3 & 1 \\
\small\texttt{ptest}  &\small\texttt{\_mm\_testz\_si128} & performs a bitwise and of two 128-bit integers; returns one, if the result is all zeros, and zero otherwise. & 1 & 1 \\
\bottomrule
\end{tabular}
\end{table}

\section{Integer compression}\label{sec:veccompression}

We consider the case
where we have lists of integers stored using 32~bits, but where the
magnitude of most integers requires fewer than 32-bits to express.
We want to compress them while spending as few CPU cycles per integer as possible.

There has been much work on the design and analysis of integer compression schemes.
Out of an earlier survey~\cite{LemireBoytsov2013decoding},
we choose 4~previously described fast compression schemes: \textsc{varint}, \textsc{S4-BP128}, \textsc{FastPFOR} and \textsc{S4-FastPFOR}. Both \textsc{S4-BP128} and \textsc{S4-FastPFOR} performed best in an exhaustive experimental comparison~\cite{LemireBoytsov2013decoding}. We review them briefly for completeness.

\subsection{\textsc{Varint}}
\label{sec:varint}

\begin{figure}\centering
\subfloat[Unsigned binary 32-bit format: for each integer, there are four unsigned byte values (in little endian representation).\vspace{1em} \label{fig:baby32}]{\begin{tikzpicture}[node distance=0cm,start chain=9 going right]
\edef\sizetape{0.2cm}    \tikzstyle{mytape}=[draw,minimum size=\sizetape]
\tikzstyle{overbrace style}=[decorate,decoration={brace,raise=2mm,amplitude=3pt}]

    \node  [on chain=9,mytape,fill=blue!20] {1};
    \node [on chain=9,mytape,fill=blue!20] {0};
    \node [on chain=9,mytape,fill=blue!20] {0};
    \node [on chain=9,mytape,fill=blue!20] {0};
    \node [on chain=9,mytape,fill=yellow!20] {0};
    \node(id9) [on chain=9,mytape,fill=yellow!20] {15};
        \node [above of=id9,node distance=0.6cm] {$4 \times 4$~bytes or 16 bytes};
    \node [on chain=9,mytape,fill=yellow!20] {0};
    \node [on chain=9,mytape,fill=yellow!20] {0};
    \node [on chain=9,mytape,fill=green!20] {1};
    \node [on chain=9,mytape,fill=green!20] {0};
    \node [on chain=9,mytape,fill=green!20] {2};
    \node [on chain=9,mytape,fill=green!20] {0};
    \node [on chain=9,mytape,fill=brown!20] {2};
    \node [on chain=9,mytape,fill=brown!20] {0};
    \node [on chain=9,mytape,fill=brown!20] {0};
    \node [on chain=9,mytape,fill=brown!20] {0};
\end{tikzpicture}
}

\subfloat[\textsc{varint} format\label{fig:babyvarint}]{%
\parbox{0.4\textwidth}
{\vspace{-5cm}\centering\begin{tikzpicture}[node distance=0cm,start chain=9 going right]
\edef\sizetape{0.2cm}    \tikzstyle{mytape}=[draw,minimum size=\sizetape,fill=blue!20]
\tikzstyle{overbrace style}=[decorate,decoration={brace,raise=2mm,amplitude=3pt}]
    \node  [on chain=9,mytape] {\textbf{1}};
    \node [on chain=9,mytape] {0};
    \node [on chain=9,mytape] {0};
    \node(id9) [on chain=9,mytape] {0};
        \node [above of=id9,node distance=0.6cm] {1 byte};
    \node [on chain=9,mytape] {0};
    \node [on chain=9,mytape] {0};
    \node [on chain=9,mytape] {0};
    \node [on chain=9,mytape] {1};
\end{tikzpicture}
\\
\begin{tikzpicture}[node distance=0cm,start chain=9 going right,start chain=10 going right]
\edef\sizetape{0.2cm}    \tikzstyle{mytape}=[draw,minimum size=\sizetape,fill=yellow!20]
\tikzstyle{overbrace style}=[decorate,decoration={brace,raise=2mm,amplitude=3pt}]
    \node(id1)  [on chain=9,mytape] {\textbf{0}};
    \node [on chain=9,mytape] {0};
    \node [on chain=9,mytape] {0};
    \node(id9) [on chain=9,mytape] {0};
        \node [above of=id9,node distance=0.6cm] {2 bytes};
    \node [on chain=9,mytape] {0};
    \node [on chain=9,mytape] {0};
    \node [on chain=9,mytape] {0};
    \node [on chain=9,mytape] {0};
    \node  [on chain=10,mytape,below =of id1] {\textbf{1}};
    \node  [on chain=10,mytape] {0};
    \node  [on chain=10,mytape] {0};
    \node  [on chain=10,mytape] {1};
    \node  [on chain=10,mytape] {1};
    \node  [on chain=10,mytape] {1};
    \node  [on chain=10,mytape] {1};
    \node  [on chain=10,mytape] {0};
\end{tikzpicture}\\
\begin{tikzpicture}[node distance=0cm,start chain=9 going right,start chain=10 going right,start chain=11 going right]
\edef\sizetape{0.2cm}    \tikzstyle{mytape}=[draw,minimum size=\sizetape,fill=green!20]
\tikzstyle{overbrace style}=[decorate,decoration={brace,raise=2mm,amplitude=3pt}]
    \node(id1)  [on chain=9,mytape] {\textbf{0}};
    \node [on chain=9,mytape] {0};
    \node [on chain=9,mytape] {0};
    \node(id9) [on chain=9,mytape] {0};
        \node [above of=id9,node distance=0.6cm] {3 bytes};
    \node [on chain=9,mytape] {0};
    \node [on chain=9,mytape] {0};
    \node [on chain=9,mytape] {0};
    \node [on chain=9,mytape] {1};
    \node(id11)  [on chain=10,mytape,below =of id1] {\textbf{0}};
    \node [on chain=10,mytape] {0};
    \node [on chain=10,mytape] {0};
    \node(id9) [on chain=10,mytape] {0};
    \node [on chain=10,mytape] {0};
    \node [on chain=10,mytape] {0};
    \node [on chain=10,mytape] {0};
    \node [on chain=10,mytape] {0};
    \node  [on chain=11,mytape,below =of id11] {\textbf{1}};
    \node [on chain=11,mytape] {0};
    \node [on chain=11,mytape] {0};
    \node(id9) [on chain=11,mytape] {0};
    \node [on chain=11,mytape] {1};
    \node [on chain=11,mytape] {0};
    \node [on chain=11,mytape] {0};
    \node [on chain=11,mytape] {0};
\end{tikzpicture}\\
\begin{tikzpicture}[node distance=0cm,start chain=9 going right,start chain=10 going right]
\edef\sizetape{0.2cm}    \tikzstyle{mytape}=[draw,minimum size=\sizetape,fill=brown!20]
\tikzstyle{overbrace style}=[decorate,decoration={brace,raise=2mm,amplitude=3pt}]
    \node(id1)  [on chain=9,mytape] {\textbf{1}};
    \node [on chain=9,mytape] {0};
    \node [on chain=9,mytape] {0};
    \node(id9) [on chain=9,mytape] {0};
        \node [above of=id9,node distance=0.6cm] {1 byte};
    \node [on chain=9,mytape] {0};
    \node [on chain=9,mytape] {0};
    \node [on chain=9,mytape] {1};

    \node [on chain=9,mytape] {0};

\end{tikzpicture}
}
}\subfloat[Packed format\label{fig:packed}]{\begin{tikzpicture}[node distance=0cm,start chain=1 going right,start chain=2 going right,start chain=3 going right,start chain=4 going right,start chain=5 going right,start chain=6 going right,start chain=7 going right,start chain=8 going right,start chain=9 going right]
\edef\sizetape{0.2cm}    \tikzstyle{mytape}=[draw,minimum size=\sizetape]
\tikzstyle{overbrace style}=[decorate,decoration={brace,raise=2mm,amplitude=3pt}]
    \node(id1)  [on chain=1,mytape,fill=blue!20] {0};
    \node [on chain=1,mytape,fill=blue!20] {0};
    \node [on chain=1,mytape,fill=blue!20] {0};
    \node [on chain=1,mytape,fill=blue!20] {0};
    \node(id9) [on chain=1,mytape,fill=blue!20] {0};
        \node [above of=id9,node distance=0.6cm] {$4\times 18$ bits or 9 bytes};
    \node [on chain=1,mytape,fill=blue!20] {0};
    \node [on chain=1,mytape,fill=blue!20] {0};

    \node [on chain=1,mytape,fill=blue!20] {1};

    \node(id21)  [on chain=2,mytape,below =of id1,fill=blue!20] {0};
    \node  [on chain=2,mytape,fill=blue!20] {0};
    \node  [on chain=2,mytape,fill=blue!20] {0};
    \node  [on chain=2,mytape,fill=blue!20] {0};
    \node  [on chain=2,mytape,fill=blue!20] {0};
    \node  [on chain=2,mytape,fill=blue!20] {0};
    \node  [on chain=2,mytape,fill=blue!20] {0};
    \node  [on chain=2,mytape,fill=blue!20] {0};

    \node(id31)  [on chain=3,mytape,below =of id21,fill=yellow!20] {0};
    \node  [on chain=3,mytape,fill=yellow!20] {0};
    \node  [on chain=3,mytape,fill=yellow!20] {0};
    \node  [on chain=3,mytape,fill=yellow!20] {0};
    \node  [on chain=3,mytape,fill=yellow!20] {0};
    \node  [on chain=3,mytape,fill=yellow!20] {0};
    \node  [on chain=3,mytape,fill=blue!20] {0};
    \node  [on chain=3,mytape,fill=blue!20] {0};

        \node(id41)  [on chain=4,mytape,below =of id31,fill=yellow!20] {0};
    \node  [on chain=4,mytape,fill=yellow!20] {0};
    \node  [on chain=4,mytape,fill=yellow!20] {1};
    \node  [on chain=4,mytape,fill=yellow!20] {1};
    \node  [on chain=4,mytape,fill=yellow!20] {1};
    \node  [on chain=4,mytape,fill=yellow!20] {1};
    \node  [on chain=4,mytape,fill=yellow!20] {0};
    \node  [on chain=4,mytape,fill=yellow!20] {0};

        \node(id51)  [on chain=5,mytape,below =of id41,fill=green!20] {0};
    \node  [on chain=5,mytape,fill=green!20] {0};
    \node  [on chain=5,mytape,fill=green!20] {0};
    \node  [on chain=5,mytape,fill=green!20] {1};
    \node  [on chain=5,mytape,fill=yellow!20] {0};
    \node  [on chain=5,mytape,fill=yellow!20] {0};
    \node  [on chain=5,mytape,fill=yellow!20] {0};
    \node  [on chain=5,mytape,fill=yellow!20] {0};

        \node(id61)  [on chain=6,mytape,below =of id51,fill=green!20] {0};
    \node  [on chain=6,mytape,fill=green!20] {0};
    \node  [on chain=6,mytape,fill=green!20] {0};
    \node  [on chain=6,mytape,fill=green!20] {0};
    \node  [on chain=6,mytape,fill=green!20] {0};
    \node  [on chain=6,mytape,fill=green!20] {0};
    \node  [on chain=6,mytape,fill=green!20] {0};
    \node  [on chain=6,mytape,fill=green!20] {0};

        \node(id71)  [on chain=7,mytape,below =of id61,fill=brown!20] {1};
    \node  [on chain=7,mytape,fill=brown!20] {0};
    \node  [on chain=7,mytape,fill=green!20] {1};
    \node  [on chain=7,mytape,fill=green!20] {0};
    \node  [on chain=7,mytape,fill=green!20] {0};
    \node  [on chain=7,mytape,fill=green!20] {0};
    \node  [on chain=7,mytape,fill=green!20] {0};
    \node  [on chain=7,mytape,fill=green!20] {0};

        \node(id81)  [on chain=8,mytape,below =of id71,fill=brown!20] {0};
    \node  [on chain=8,mytape,fill=brown!20] {0};
    \node  [on chain=8,mytape,fill=brown!20] {0};
    \node  [on chain=8,mytape,fill=brown!20] {0};
    \node  [on chain=8,mytape,fill=brown!20] {0};
    \node  [on chain=8,mytape,fill=brown!20] {0};
    \node  [on chain=8,mytape,fill=brown!20] {0};
    \node  [on chain=8,mytape,fill=brown!20] {0};

        \node  [on chain=,mytape,below =of id81,fill=brown!20] {0};
    \node  [on chain=9,mytape,fill=brown!20] {0};
    \node  [on chain=9,mytape,fill=brown!20] {0};
    \node  [on chain=9,mytape,fill=brown!20] {0};
    \node  [on chain=9,mytape,fill=brown!20] {0};
    \node  [on chain=9,mytape,fill=brown!20] {0};
    \node  [on chain=9,mytape,fill=brown!20] {0};
    \node  [on chain=9,mytape,fill=brown!20] {0};

\end{tikzpicture}

}
\caption{The sequence of numbers  1, 3840, 131073 and 2 using 3 different data formats.
We use a left to right, top to bottom representation, putting most significant bits first in each byte.\label{fig:babyexample}}
\end{figure}
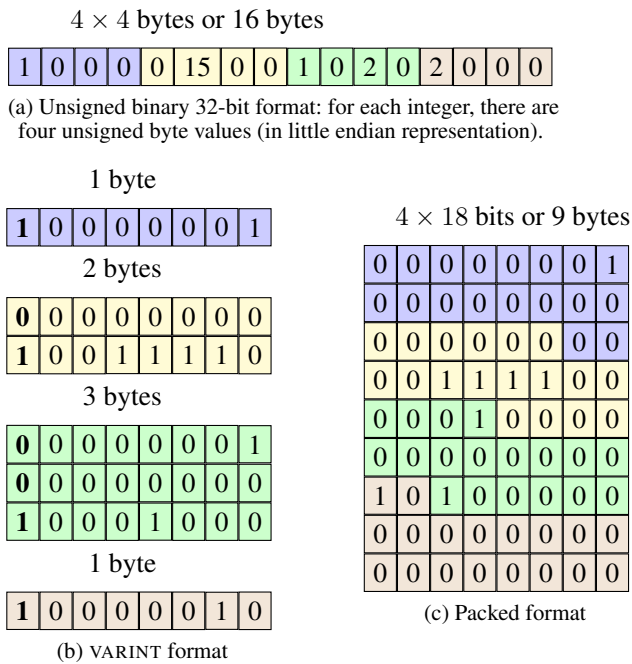

 Many authors such as Culpepper and Moffat~\cite{Culpepper:2010:ESI:1877766.1877767} use variable byte codes (henceforth \textsc{varint}) also known as escaping~\cite{Transier:2010:EBA:1877766.1877768} for compressing integers. It was first described by Thiel and Heaps~\cite{thiel1972program}. For example, we might code integers in $[0,2^7)$ using a single byte, integers in $[2^7, 2^{14})$ using two bytes and so on.
As an example, consider the integers 1, 3840, 131073, and 2, and Fig.~\ref{fig:babyexample}. In Fig.~\ref{fig:baby32}, we give the usual 32-bit integer layout for these integers, it uses 16 bytes. In binary format, these numbers can be written 1,
 111100000000, 100000000000000001 and 10 (writing the most significant bits first). The first integer can be written using a single byte since it is in $[0,2^7)$. To indicate that the byte corresponds to a complete integer, we set the most significant bit to 1: 10000001. The integer 3840 is in  $[2^7, 2^{14})$, so we represent it using two bytes. The first byte corresponds to the first 7~bits and has 0  as its most significant bit to indicate that it does not represent a full integer (000000001). The second byte includes the remaining bits and has 1 as its most significant bit (10011110). We proceed similarly for the integers 131073 and 2 (see Fig.~\ref{fig:babyvarint}). Overall, \textsc{varint} using 7~bytes instead of the original 16~bytes.
 \textsc{Varint} does not always compress well: it always uses at least one byte per integer. However, if most integers can be represented with a single byte, then it offers competitive decompression speed~\cite{LemireBoytsov2013decoding}.
Stepanov  et al.~\cite{Stepanov:2011:SDP:2063576.2063627} reviewed \textsc{varint}-like alternatives that use SIMD instructions. Lemire and Boytsov~\cite{LemireBoytsov2013decoding} found that the fastest such alternative (varint-G8IU) was slower and did not compress as well as other SIMD-based schemes.

\subsection{Bit packing and unpacking}

\label{sec:bitpackingonly}

%\daniel{We now have a separate section for bit packing and unpacking.}
%ok

Consider once more the integers 1, 3840, 131073, and 2, and Fig.~\ref{fig:babyexample}. The largest of them is 131073 and it can be represented using 18~bits in binary format (as 100000000000000001). Thus we could decide to write
the four integers using exactly 18~bits per integer. The result in memory might look like Fig.~\ref{fig:packed} and span 9~bytes.
 We call this process \emph{bit packing}, and the reverse process \emph{bit
unpacking} (see Table~\ref{ref:termino} for a summary of our terminology).

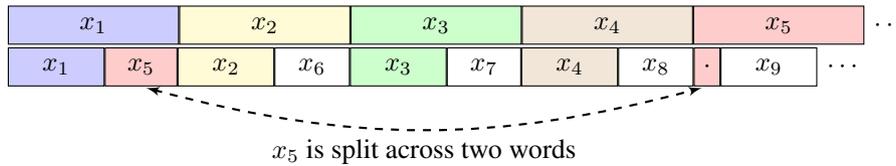
\begin{figure}\centering
\parbox{0.8\textwidth}{%
\begin{tikzpicture}[node distance=0cm,start chain=9 going right]
\edef\sizetape{0.07cm}    \tikzstyle{mytape}=[draw,minimum width=32*\sizetape,minimum height=0.5cm]
\tikzstyle{overbrace style}=[decorate,decoration={brace,raise=2mm,amplitude=3pt}]
    \node  [on chain=9,mytape,fill=blue!20] {$x_1$};
    \node [on chain=9,mytape,fill=yellow!20] {$x_2$};
    \node [on chain=9,mytape,fill=green!20] {$x_3$};
    \node [on chain=9,mytape,fill=brown!20] {$x_4$};
    \node [on chain=9,mytape,fill=red!20] {$x_5$};
    \node(id9) [on chain=9,mytape,minimum width=10*\sizetape,draw=none] {$\cdots$};
\end{tikzpicture}

\begin{tikzpicture}[node distance=0cm,start chain=9 going right]
\edef\sizetape{0.07cm}    \tikzstyle{mytape}=[draw,minimum height=0.5cm]
\tikzstyle{overbrace style}=[decorate,decoration={brace,raise=2mm,amplitude=3pt}]
    \node  [on chain=9,mytape,minimum width=18*\sizetape,fill=blue!20] {$x_1$};
    \node(x51) [on chain=9,mytape,minimum width=13.5*\sizetape,fill=red!20] {$x_5$};
    \node [on chain=9,mytape,minimum width=18*\sizetape,fill=yellow!20] {$x_2$};
    \node [on chain=9,mytape,minimum width=14*\sizetape] {$x_6$};
    \node [on chain=9,mytape,minimum width=18*\sizetape,fill=green!20] {$x_3$};
    \node [on chain=9,mytape,minimum width=13.8*\sizetape] {$x_7$};
    \node [on chain=9,mytape,minimum width=18*\sizetape,fill=brown!20] {$x_4$};
    \node [on chain=9,mytape,minimum width=14*\sizetape] {$x_8$};

    \node(x52)  [on chain=9,mytape,minimum width=4*\sizetape,fill=red!20] {$\cdot$};
    \node [on chain=9,mytape,minimum width=18*\sizetape] {$x_{9}$};
    \node [on chain=9,mytape,minimum width=10*\sizetape,draw=none] {$\cdots$};

\draw[<->, >=latex', shorten >=2pt, shorten <=2pt, bend right=15, thick, dashed]
    (x51.south) to node[auto, swap] {$x_5$ is split across two words}(x52.south);

\end{tikzpicture}
}
\caption{\label{fig:interleaved} Packing 32-bit integers $x_1, x_2,\ldots$ (top) to 18~bits per integer (bottom)
using the interleaved packed format for SIMD processing}
\end{figure}

The consecutive layout of Fig.~\ref{fig:babyexample} is adequate for scalar processing: unpacking an integer can be done efficiently using shifts and bitwise logical operations. However, when using SIMD processing, we want to pack and unpack several  (e.g., 4) integers at once.
For this reason, we choose to interleave the packed integers as in
Fig.~\ref{fig:interleaved}. In this example, the integers $x_1, x_2, x_3, x_4$ are packed to the first 18~bits  of 4 consecutive 32-bit words. The first 14~bits of the next integers $x_5, x_6, x_7, x_8$ are packed to the next 14~bits of these 32-bit integers,
%\leoinline{isn't the other way around, the first bits of the next four ... are packed to. not quite sure:}
%\danielinline{I think this was correct but I have rephrased so as to be more explicit. I think it is impossible to understand from the prose alone in any case, the reader needs to look at the picture.}
%% I think now it's easier to understand.
and the remaining $18-14=4$~bits are packed to the first bits of the next four consecutive 32-bit words.
See Lemire and Boytsov~\cite{LemireBoytsov2013decoding} for a more detailed discussion and a comparison with an alternative layout.

Only 4~basic operations are required for bit unpacking:  bitwise or,
bitwise and, logical shift right, and logical shift left.  The
corresponding 128-bit SSE2 instructions operating on packed 32-bit integers
 are \texttt{por}, \texttt{pand},  \texttt{psrld}, and \texttt{pslld} (in Intel and AMD processors).
 For a given bit width $b$, no branching is required for bit packing or bit unpacking. Thus, we can create one bit unpacking function for each bit width $b$ and select the desired one using
 an array of function pointers or a \texttt{switch/case} statement.
In the scalar case, it is most convenient to pack and unpack integers in units of 32~integers. Given a bit width $b$, the
%\leo{unpacking procedures outputs...?}
%%% Daniel: good catch.
unpacking procedure outputs $b$~32-bit integers. In the vectorized case, we pack and unpack integers in blocks of 128~integers, so that the
%\leo{Again, this should probably be better to say unpacking procedure...}
%% Daniel: another good catch.
%% Ok
unpacking procedure--corresponding to bit width $b$--generates $b$ 128-bit vectors.
A generic bit unpacking procedure for a block of 128~integers is given by Algorithm~\ref{algo:unpacking} and discussed again in~\S~\ref{sec:diffcod}.

%The number 128 was  chosen to match the bit width
%of the 4-integer vector based on the following insight.  Suppose you
%encode 128~integers using exactly $b$~bits per integer.  The result will
%use $128 \times b$~bits of memory ($b \times 16$~bytes of memory).  In this way, our
%algorithm can always make full use of 128-bit vector processing on
%each 16-byte chunk, with larger bit widths simply requiring more
%chunks.

\subsection{\textsc{S4-BP128}}

\label{sec:s4bp128}

Bit packing suggests a simple  scheme: regroup the integers into blocks (e.g. 128 integers) and pack them as concisely as possible, while recording the bit width (e.g., 18~bits per integer) using an extra byte.
We call this approach \textsc{binary packing}.
Binary packing is closely related to Frame-Of-Reference (FOR)~\cite{655800}
and it has been called
PackedBinary~\cite{Anh:2010:ICU:1712666.1712668}.

Binary packing can be fast.
Indeed, Catena et al.~\cite{Catena2014} found that it offered the
best speed in a search engine setting.
Our fastest family of compression schemes is an instance of binary packing:
\textsc{S4-BP128}. The ``S4'' stands for 4-integer SIMD, ``BP'' stands for ``Binary
Packing'', and ``128'' indicates the number of integers encoded in each
block.
%The number 128 was  chosen to match the bit width
%of the 4-integer vector based on the following insight.  Suppose you
%encode 128~integers using exactly $b$~bits per integer.  The result will
%use $128 \times b$~bits of memory ($b \times 16$~bytes of memory).  In this way, our
%algorithm can always make full use of 128-bit vector processing on
%each 16-byte chunk, with larger bit widths simply requiring more
%chunks.

In the \textsc{S4-BP128} format, we decompose arrays into meta-blocks of 2048~integers, each containing 16~blocks of 128~integers. Before bit packing the 16~blocks, we write 16~bit widths ($b$) using one byte each. For each block, the bit width is the smallest value $b$ such that all corresponding integers are smaller than $2^b$. The value $b$ can range from 0 to 32. In practice, lists of integers are rarely divisible by 2048. We handle remaining blocks of 128~integers separately. For each such block, we write a byte containing a bit width $b$, followed by the corresponding 128~integers in bit packed form. Finally, we compress the remaining integers (less than 128~integers) using  \textsc{varint}.
%\leoinline{This is not clear, do you mean that the remainder contains less than 2048 integers and all these integers are compressed using \textsc{varint}?}
%\danielinline{No, it should be clear that we always compress fewer than 128 integers using varint. See revised text where I tried to be clearer.}
% Leo: now it is.

\subsection{\textsc{FastPFOR}, \textsc{SIMD-FastPFOR}, and \textsc{S4-FastPFOR}}
\label{sec:fastpfor}

The downside of the \textsc{S4-BP128} approach is that the largest integers in a block of 128~integers determine the compression ratio of all these integers. We could use smaller blocks (e.g., 32) to improve compression. However, a better approach for performance might be \emph{patching}~\cite{1617427} wherein the block is first decompressed using a smaller bit width,
and then a limited number of entries requiring greater bit widths are
overwritten (``patched'') using additional information.
That is, instead of picking the bit width $b$ such that all integers are smaller than $2^b$, we pick a different bit width $b'$ that might be smaller than  $b$.  That is, we only bit pack the least significant $b'$-bits from each integer.  We must still encode the missing information for  all integers larger than or equal to $2^{b'}$: we call each such integer an exception.

In our version of patched coding (\textsc{FastPFOR}, for ``fast patched frame-of-reference''), we proceed as in \textsc{S4-BP128}, that is we bit pack blocks of 128~integers.
To complement this data, we use a \emph{metadata} byte array.
For each block, we store both bit widths $b$ and $b'$, as well as the number of exceptions.
Each  location
where an exception should be applied is also stored using one byte in the \emph{metadata} byte array. It remains to store the most significant $b-b'$~bits of the integers larger than or equal to $2^{b'}$. We collect these exceptions over many blocks (up to 512~blocks) and then bit pack them together, into up to 32~bit packed arrays (one for each possible value of $b-b'$ excluding zero). For speed, these arrays are padded up to a multiple of 32~integers. The final data format is an aggregation of the bit packed blocks (using $b'$~bits per integer), the \emph{metadata} byte array and the bit packed arrays corresponding to the exceptions.

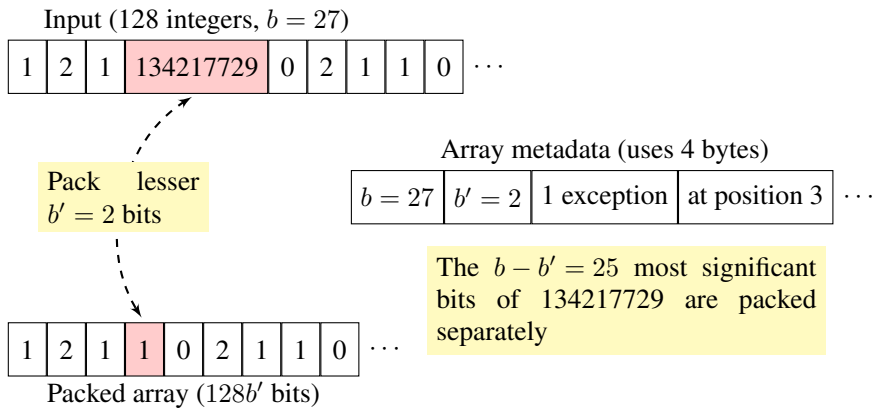
\begin{figure}
\centering
\parbox{0.8\textwidth}{%
\begin{tikzpicture}[node distance=0cm,start chain=9 going right,start chain=13 going right,start chain=15 going right,start chain=17 going right]
\edef\sizetape{0.07cm}    \tikzstyle{mytape}=[draw,minimum height=0.7cm,,minimum width=0.5cm]
\tikzstyle{overbrace style}=[decorate,decoration={brace,raise=2mm,amplitude=3pt}]
    \node(B1) [on chain=9,mytape] {1};
    \node [on chain=9,mytape] {2};
    \node [on chain=9,mytape] {1};
    \node(O1) [on chain=9,mytape,fill=red!20] {134217729};
    \node [on chain=9,mytape] {0};
    \node [on chain=9,mytape] {2};
    \node [on chain=9,mytape] {1};
    \node [on chain=9,mytape] {1};
    \node [on chain=9,mytape] {0};
    \node [on chain=9,mytape,draw=none] {$\cdots$};

            \node [above of=O1,node distance=0.6cm] {Input (128 integers, $b=27$)};

    \node [on chain=13,mytape,below=3cm of B1] {1};
    \node [on chain=13,mytape] {2};
    \node [on chain=13,mytape] {1};
    \node(E1) [on chain=13,mytape,fill=red!20] {1};
    \node(Z1) [on chain=13,mytape] {0};
    \node [on chain=13,mytape] {2};
    \node [on chain=13,mytape] {1};
    \node [on chain=13,mytape] {1};
    \node [on chain=13,mytape] {0};
    \node [on chain=13,mytape,draw=none] {$\cdots$};

\node[below right=2cm and 5cm of B1,fill=yellow!30]{\parbox{5cm}{The $b-b'=25$ most significant bits of 134217729 are packed separately}};

    \node [on chain=15,mytape,below right=1cm and 4cm of B1] {$b=27$};
    \node [on chain=15,mytape] {$b'=2$};
    \node(meta1) [on chain=15,mytape] {1 exception};
    \node [on chain=15,mytape] {at position 3};
    \node [on chain=15,mytape,draw=none] {$\cdots$};

            \node [above of=meta1,node distance=0.6cm] {Array metadata (uses 4 bytes)};

            \node [below of=Z1,node distance=0.6cm] {Packed array ($128 b'$~bits)};

\draw[<->, >=latex', shorten >=2pt, shorten <=2pt, thick, bend right=45,dashed]
    (O1.south) to node[fill=yellow!30] {\parbox{2cm}{Pack lesser $b'=2$~bits}}
    (E1.north);
\end{tikzpicture}
}
\caption{Example of compression using \textsc{FastPFOR} (and \textsc{S4-FastPFOR}). The data corresponding to a block of 128~integers is located in a packed array (using $b'$~bit per integer, in a metadata byte array and, finally, in a corresponding packed array of exceptions.  \label{fig:fastpfor}}
\end{figure}

As an illustration, consider Fig.~\ref{fig:fastpfor}. We consider a block of 128~integers beginning with 1,2, 1, 134217729, 0. We assume that all values in the block except for 134217729 do not exceed three and, thus, can be encooded using two bits each. In this case, we have that $b=27$ and binary packing would thus require 27~bits per integer---even if most of the integers could fit in two bits. With \textsc{FastPFOR}, we might set $b'=2$ and pack only the least significant two bits of each integer. To allow decoding, we have to record that there is one exception at location 3 as well as the values of $b$ and $b'$. Each of these numbers can be stored using one byte. We are left to code the $b-b'=25$~most significant bits of 134217729. We can store these bits in an array to be packed later. Once we have processed many blocks of 128~integers, there might be several similar exceptions that need the storage of their 25~most significant bits: they are stored and packed together.  During decompression, these most significant bits can be unpacked as needed from one of 32~different arrays (one for each possible value of $b-b'$ excluding zero).

Decompression using \textsc{FastPFOR} is similar to decompression using  \textsc{S4-BP128}, except that, after bit unpacking the least significant $b'$~bits for the integers of a block, we must proceed with the \emph{patching}. For this purpose, the packed arrays of exception values are unpacked as needed and we loop over the exceptions.

The integer $b$ is determined by the data---as the number of bits required to store the largest integer in the block. However, we are free to set $b'$ to any integer in $[0,b]$. For each block, we pick $b'$ to minimize $128 \times b' + c(b') (b-b' + 8)$ where $c(b')$ is the number exceptions generated for a given value $b'$ (e.g., $c(b)=0$). This heuristic
%\leoinline{Should we say that this is actually a lower bound for the number of bits (minus a few constants) that is close to an actual number of bits under the assumption that there is little waste in encoding the exceptions, i.e., that we have a lot of them for each bit value?}
%\danielinline{ I say no. Sometimes, it is best to take the formula as it is and not over-explain things. The reader is given a formula. It fully defines how the codec works. That should be good enough for this paper.}
% ok, I didn't have a strong opinion about this.
 for picking $b'$ can be computed quickly if we first
 tabulate how many integers in the block are less than $2^x$ for $x=0, \ldots, b$. For this purpose, we use the assembly instruction \texttt{bsr}
(Bit Scan Reverse) to calculate the $\log_2$ of each integer.

We can compare \textsc{FastPFOR} with other patched schemes based on Lemire and Boytsov's results~\cite{LemireBoytsov2013decoding}: the original scheme from Zukowski et al.~\cite{1617427} is faster than \textsc{FastPFOR} but
has worse compression ratios (even worse than the bit packing method \textsc{S4-BP128} that does not use patching),
OptPFD~\cite{yan2009inverted} by Yan el al.\ sometimes compresses better than \textsc{FastPFOR},
but can be up to two~times slower, and their NewPFD~\cite{yan2009inverted} is typically slower and offers worse compression ratios than \textsc{FastPFOR}.
A format inspired by \textsc{FastPFOR} (where bits from  exceptions are bit packed) is a part of the search engine Apache Lucene as of version~4.5.\footnote{\url{http://lucene.apache.org/core/4_9_0/core/org/apache/lucene/util/PForDeltaDocIdSet.html}}
In contrast, Catena et al.~\cite{Catena2014} found  that
\textsc{FastPFOR} provided no response time benefit compared to NewPFD and OptPFD when compressing document identifiers.
 Possibly, the discrepancy can be
explained by the fact that they divided their arrays into
small chunks of 1024~integers prior to compression. Indeed,
\textsc{FastPFOR} is most effective when it is allowed to
aggregate the exceptions over many blocks of 128~integers.

Because \textsc{FastPFOR} relies essentially on bit packing for compression, it is easy to vectorize it using the same SIMD-based bit packing and unpacking functions used by \textsc{S4-BP128}. We call the resulting scheme \textsc{S4-FastPFOR}. In the original  scheme   by Lemire and Boytsov~\cite{LemireBoytsov2013decoding} (called \textsc{SIMD-FastPFOR}), the bit-packed exceptions were padded up to multiples of 128~integers instead of multiples of 32~integers as in \textsc{FastPFOR}. While this insured that all memory pointers were aligned on 16~bytes boundaries, it also adversely affected compression. Our \textsc{S4-FastPFOR}  has essentially the same data format as \textsc{FastPFOR} and, thus, the same compression ratios. These changes require that we use  \begin{inparaenum}[(1)]\item scalar bit packing/unpacking for up to 32~integers per packed array and \item  unaligned load and store SIMD instructions. \end{inparaenum}
Neither of these differences significantly impacts performance, and the
compression ratios improve by $\approx$\SI{5}{\percent}.

\begin{table}
\caption{Some of our terminology\label{ref:termino}}\centering
\begin{tabular}{p{1.5in}p{3in}}
\toprule
Differential coding & Strategy to transform lists of sorted integers into lists of small integers by computing differences between nearby, but not necessarily adjacent, integers. We consider four types of differential coding (\textsc{D1}, \textsc{D2}, \textsc{DM}, \textsc{D4}). To recover the initial array, it is necessary to compute a prefix sum or its analog.
%\leoinline{Maybe it should be explained somehow better. Yet, we do not exactly compute a prefix sum in the case of D2, DM and D4.}

%\danielinline{One referee was confused that we used unpacking to refer to something that included differential coding:}
Optionally, we can integrate differential coding in the packing and unpacking. In such a case, unpacking includes the computation of the prefix sum.\\[1ex]
pack (unpack) & To encode (resp.\ decode) 32-bit integers to (resp.\ from) data blocks where all integers use a fixed number of bits (bit width). Packing and unpacking routines are used as part of integer compression algorithms such as binary packing. \\[1ex]
binary packing & Compression scheme that packs blocks of integers (e.g., 128 integers) using as few bits as possible (e.g., the \textsc{S4-BP128} family)\\[1ex]
patched coding & Compression strategy that uses binary packing to encode most of the data, while also encoding exceptions separately (e.g., the \textsc{FastPFOR} scheme) to improve the compression ratio. %\\[1ex]
%\textsc{S4-} & Prefix used to indicate that we use SIMD instructions processing 4~integers at a time
\\
\bottomrule
\end{tabular}
\end{table}

\section{Differential coding}\label{sec:diffcod}

Differential coding takes a sorted integer list $x_1, x_2, \ldots $ and replaces it
with successive differences (or deltas) $\delta_2, \delta_3, \ldots = x_2-x_1, x_3-x_2, \ldots$. If we keep the first value intact ($x_1$), this transformation is invertible.
We can then apply various integer compression schemes on the deltas.
 Since the deltas are smaller than the
original integers in the list, we get
better compression ratios.

If the sorted lists contain no repetitions, we can further subtract one from all deltas as we know
that they are greater than zero.
However, this only offers a significant benefit in compression if
we expect many of the deltas to be close to zero. And, in such cases, other schemes such as bit vectors might be more appropriate. Thus, we do not consider this option any further.

Computing deltas during compression is an
inexpensive operation that can be easily accelerated with superscalar execution or even SIMD instructions. However, recovering the original list from the deltas when decompressing can be more time
consuming because of the inherent data-dependencies:  the value of
each recovered integer can only be calculated after the value of the
integer immediately preceding it is known.
 Indeed, it involves the computation of a prefix sum: $x_i = x_{i-1} + \delta_i$.
A naive implementation could end up using one or more CPU cycles per
integer just to calculate the prefix sum.
For a moderately fast scheme such as  \textsc{varint}, this is not a concern, but for faster
schemes, computation of the prefix sum can become a performance bottleneck.
To alleviate this problem, we can sacrifice some compressibility, by computing the deltas on a four-by-four basis (henceforth \textsc{D4},
because it compares index differences of 4): $\delta_i = x_i - x_{i-4}$~\cite{LemireBoytsov2013decoding}.
 Although fast, this approach also generates larger deltas. %\daniel{referee wants us to explain our acronyms.}

As further refinements, we  consider four~different forms of differential coding that offer different compression/speed trade-offs. Each form has a corresponding inverse---for simplicity, we use the term \emph{prefix sum} to refer to all these inverse functions. As before, we assume 128-bit vectors and process 32-bit integers.

\begin{itemize}[leftmargin=*]
\item The fastest is \textsc{D4} which computes the deltas four-by-four:  e.g.,
$(\delta_5,\delta_6,\delta_7,\delta_8) = (x_5,x_6,x_7,x_8) -  (x_1,x_2,x_3,x_4)$.
 We expect the deltas to be $4\times$ larger, which degrades the compression by approximately
two bits per integer.  However,  a single 1-cycle-latency SIMD
instruction (\texttt{paddd} in SSE2) can correctly calculate the prefix sum of
four consecutive integers.
\item The second fastest is \textsc{DM}. It is similar to \textsc{D4} except that instead of subtracting the previously decoded vector of integers, we subtract only the largest of these  integers: $\delta_{4 i+j} = x_{4 i+j} -  x_{4 i - 1}$. E.g., \begin{align*}(\delta_5,\delta_6,\delta_7,\delta_8) = (x_5,x_6,x_7,x_8) -  (x_4,x_4,x_4,x_4).\end{align*}  We expect the deltas to be $2.5 \times$ larger on average. Compared to the computation of the prefix sum with \textsc{D4}, \textsc{DM} requires one extra instruction (\texttt{pshufd} in SSE2) to copy the last component to all components:  $(x_1,\ldots,x_4) \to (x_4,\ldots,x_4)$. On Intel processors, the \texttt{pshufd} instruction is fast.
\item The third fastest is \textsc{D2}: $\delta_i = x_i - x_{i-2}$. E.g., \begin{align*}(\delta_5,\delta_6,\delta_7,\delta_8) = (x_5,x_6,x_7,x_8) -  (x_3,x_4,x_5,x_6).\end{align*} The deltas should be only $2 \times$~larger on average. The prefix sum for \textsc{D2} can be implemented using 4~SIMD instructions.
\begin{enumerate}
\item Shift the delta vector by 2 integers (in SSE2 using \texttt{pslldq}): e.g.,
$(\delta_5,\delta_6,\delta_7,\delta_8) \to  (0,0,\delta_5,\delta_6)$.
\item Add the original delta vector with the shifted version: e.g.,
$(\delta_5,\delta_6,\delta_5+\delta_7,\delta_6+\delta_8)$.
\item Select from the previous vector the last two integers and copy them twice (in SSE2 using \texttt{pshufd}), e.g.,
$(x_1,x_2,x_3,x_4) \to (x_3,x_4,x_3,x_4)$.
\item Add the results of the last two operations.
\end{enumerate}
\item The slowest approach is \textsc{D1} which is just the regular differential coding ($\delta_i=x_i-x_{i-1}$). It generates the smallest deltas.
We compute it with a well-known approach using 6 SIMD instructions.
\begin{enumerate}
\item The first two steps are as with \textsc{D2} to generate $(\delta_5,\delta_6,\delta_5+\delta_7,\delta_6+\delta_8)$. Then we take this result, shift it by one integer and add it to itself. Thus, we get:
\begin{align*}
(\delta_5,\delta_5+\delta_6,\delta_6+\delta_5+\delta_7,\delta_5+\delta_6+\delta_7+\delta_8).
\end{align*}
\item We copy the last integer of the previous vector to all components of a new vector. Thus, we generate
$(x_4,x_4,x_4,x_4)$.
\item We add the last two results to get the final answer.
\end{enumerate}
\end{itemize}

We summarize the different techniques (\textsc{D1}, \textsc{D2}, \textsc{DM}, \textsc{D4}) in Table~\ref{table:compd}. We stress that the number of instructions is not a measure of running-time performance, if only because of superscalar execution.
Our analysis is for 4-integer SIMD instructions: for wider SIMD instructions,
the number of instructions per integer is smaller.
%\leo{Probably, we should not talk
%about quickly diminishing number of instructions per integer as if we have asymptotic behavior here.
%It takes years to double registers' width.}
%\daniel{The current text does not say that we have
%quickly diminishing number of instructions per integer.}
% Leo: ok

\begin{table}
\caption{Comparison between the 4~vectorized differential coding techniques with 4-integer SIMD instructions\label{table:compd}}\centering
\begin{tabular}{lcS}
              &   \multicolumn{1}{c}{size of deltas} & \multicolumn{1}{c}{instructions/int} \\ \hline
\textsc{D1}   & $1.0 \times$ & 1.5 \\
\textsc{D2}  & $2.0 \times$ &  1 \\
\textsc{DM}  & $2.5 \times$ & 0.5 \\
\textsc{D4}  & $4.0 \times$ & 0.25 \\
\end{tabular}
\end{table}

 Combining a compression scheme like \textsc{varint} and differential coding is not a problem.
Since we decode integers one at a time,
we can easily integrate the computation of the prefix sum into the decompression function:
as soon as a new integer is decoded, it is added to the previous integer.
%\leoinline{This is not

%exactly accurate as we use non-standard differential encodings.}\danielinline{No. For \textsc{varint}, we use standard (scalar) differential coding. This paragraph is accurate.}
% In the previous paper we also experiment with D4, but, anyways, this is minor.

For \textsc{S4-BP128} and \textsc{S4-FastPFOR}, we could integrate differential coding at the block level. That is,  we could  unpack 128~deltas  and then calculate the prefix sum to convert
these deltas back to the original integers. Though the decompression requires two passes over the same small block of integers, it is unlikely to cause many expensive cache misses.

Maybe surprisingly,  we can do substantially better, at least for schemes such as \textsc{S4-BP128}. Instead of using two passes, we can use a single pass where we do  both the bit unpacking and the computation of the prefix sum.  In some cases, the one-pass approach is almost twice as fast as the  two-pass approach.

Indeed, when unpacking a block of 128~integers, we store 32~SSE registers in memory. Such store operations have a limited throughput of one per cycle on recent Intel processors~\cite{fog2014instruction}.
However, if the prefix sum computation is a separate step, we need to reload the recently unpacked data, compute the prefix sum and then store the data again. In effect, we end up having to store 64~SSE registers per block of 128~integers. Hence,  we need  at least 0.5~cycles to unpack an integer if the prefix sum computation is separate. These store operations can become a bottleneck. A two-step unpacking has a theoretical speed limit of 7.2~billion 32-bit integers per second on a \SI{3.60}{\GHz} processor.
Yet  we can nearly reach 9~billion 32-bit integers per second with a one-step unpacking (see \S~\ref{sec:expbitunpacking}).
%\daniel{New explanation. It is not superscalarity, stupid.}
%\leo{Leo, I am not sure, but I like the explanation, do we want to backup the latency numbers?
%perhaps, Agner Fog Instruction tables? I don't know how interpret Agner's numbers, though.}

Henceforth we use the term unpacking to also refer to the process where we both unpack and compute the prefix sum.
Algorithm~\ref{algo:unpacking} illustrates the bit unpacking routine for a block of 128~integers. It takes as a parameter
a SIMD prefix-sum function $P$ used at lines~\ref{line:ps1} and~\ref{line:ps2}:  for \textsc{D4}, we have $P(t,v)=t+v$ (an element-wise addition of two vectors). Omitting lines~\ref{line:ps1} and~\ref{line:ps2} disables differential coding. In practice,
we generate one such function for each bit width $b$ and for each prefix-sum function $P$.
The prefix sum always starts from an initial vector ($v$). We set $v=(0,0,0,0)$ initially and then, after decompressing each block of 128~integers, $v$ is set to the last 4~integers decoded.

Beside the integration of differential coding with the bit unpacking, we have also improved
over  Lemire and Boytsov's bit unpacking~\cite[Fig.~7]{LemireBoytsov2013decoding}
in another way:
whereas each of their procedures may require several masks, our
implementation uses a single mask per procedure (see line~\ref{line:mask} in Algorithm~\ref{algo:unpacking}). Given that we only have 16~SIMD registers on our Intel processors, attempting to keep several of them occupied with constant masks can be wasteful.

Unfortunately, for \textsc{S4-FastPFOR},
it is not clear how to integrate bit unpacking and computation of the prefix sum. Indeed, \textsc{S4-FastPFOR} requires three separate operations in sequence: bit unpacking, patching and computing the prefix sum. Patching must happen after bit unpacking, but before the prefix  sum. This makes tight integration between patching and the prefix sum difficult: we tried various approaches
but they did not result in performance improvement.

\begin{algorithm}
\caption{\label{algo:unpacking}Unpacking procedure using 128-bit vectors with integrated differential coding. We write  $\ShiftRight$ for the bitwise \emph{zero-fill} right shift, $\ShiftLeft$ for the bitwise left shift, $\BitAnd$ for the bitwise AND, and  $\BitOr$ for the bitwise OR\@.
The binary function $P$ depends on the type of differential coding (e.g., to disable differential coding set  $P(t,v)=t$).
 }
\centering
\begin{algorithmic}[1]
\STATE \textbf{input}:  a bit width $b$, a list of 32-bit integers $y_1, y_2,\ldots,y_{b}$, prefix-sum seed vector $v$
\STATE \textbf{output}: list of 128~32-bit integers in $[0,2^b)$
\STATE $w \leftarrow $ empty list
\STATE $M \leftarrow (2^b-1,2^b-1,2^b-1,2^b-1)$ \COMMENT{Reusable mask} \label{line:mask}
\STATE $i \leftarrow 0$
\FOR{$k = 0,1,\ldots, b-1$}
\WHILE{$i+b \leq 32 $}
\STATE $t\leftarrow (y_{1+4k} \ShiftRight i, y_{2+4k}  \ShiftRight i, y_{3+4k}  \ShiftRight i, y_{4+4k}  \ShiftRight i)$
\STATE $t\leftarrow t \BitAnd M $  \COMMENT{Bitwise AND with mask}
\STATE $t \leftarrow P(t,v)$ and $v \leftarrow t$ \COMMENT{Prefix sum}\label{line:ps1}
\STATE append integers $t_1, t_2, t_3, t_4$ to list $w$
\STATE $i \leftarrow i +b$
\ENDWHILE
\IF{$i < 32$}
\STATE $\begin{aligned}z\leftarrow &  (y_{5+4k} \ShiftLeft 32-i, y_{6+4k}  \ShiftLeft 32-i,\\
& y_{7+4k}  \ShiftLeft 32-i, y_{8+4k}  \ShiftLeft 32-i) \BitAnd M\end{aligned}$
\STATE $\begin{aligned}t\leftarrow & (y_{1+4k} \ShiftRight i, y_{2+4k}  \ShiftRight i,\\
&  y_{3+4k}  \ShiftRight i, y_{4+4k}  \ShiftRight i)
\BitOr z\end{aligned}$
\STATE $t \leftarrow P(t,v)$ and $v \leftarrow t$ \COMMENT{Prefix sum} \label{line:ps2}
\STATE append integers $t_1, t_2, t_3, t_4$ to list $w$
\STATE $i \leftarrow i + b - 32$
\ELSE
\STATE $i \leftarrow 0$
\ENDIF
\ENDFOR
\RETURN $w$
\end{algorithmic}
\end{algorithm}

\section{Fast intersections}\label{sec:fastintersection}

Consider lists of uncompressed  integers.
To compute the intersection between several sorted lists quickly, a competitive approach is Set-vs-Set (SvS): we sort the lists in the order of  non-decreasing cardinality and intersect them two-by-two, starting with the smallest.
A textbook intersection algorithm between two lists (akin to the merge sort algorithm) runs in time $O(m+n)$ where the lists have length $m$ and $n$ (henceforth we call it \textsc{scalar}). Though it is competitive when $m$ and $n$ are similar, there are better alternatives when $n \gg m$.
Such alternative algorithms assume that we  intersect a small list  $r$ with  a large list $f$.
They iterate over the small list: for each element $r_i$, they seek a match $f_j$ in the second list using some search procedure. Whether there is a match or not, they advance in the second list up to the first point where the value is at least as large as the one in the small list ($f_j \geq r_i$).  The search procedure assumes that the lists are sorted so it can skip values.
A popular algorithm uses  galloping search~\cite{bentley1976almost} (also called exponential search): we pick the next available integer $r_i$ from the small list and seek an integer at least as large in the other list, looking first at the next available value, then looking twice as far, and so on.
We keep doubling the distance until we find the first integer $f_j \ge r_i$ that is
not smaller than $r_i$.
Then, we use binary search to advance in the second list to the first value larger than or equal to $r_i$.
The binary  search range is from the current position to the position of value $f_j$.
Computing galloping intersections requires only $O(m \log n)$ time which is better than $O(m+n)$ when $n \gg m$.

\subsection{Existing SIMD intersection algorithms}\label{sec:existing}
%The vectorization of algorithms and data structures has a long history dating back to the 1960s~\cite{1447203}.
%\leoinline{
%It looks very strange that given
%the history of vectorization only two algorithms are available.
%I know that the "history" is actually more discussing then implementing.
%In addition, perhaps, something was implemented by for large machines, not commodity PCs.
%These machines may have had very exotic instructions not available on Intel.
%So, I suggest that we either delete this old reference from the sixties,
%or write something like:
% ``Yet, we could find only two algorithms ... designed for Intel processors.''
%}
%%%%%%%%% Daniel: Ok, deleting this reference to the 60s.
Though there is much related work~\cite{Zhou:2002:IDO:564691.564709},
we could find only two algorithms for the
 computation of intersections over sorted
% \leo{Over sorted \textbf{uncompressed}?}
% Daniel: Ok, Made as clear as possible.
 uncompressed lists of integers using SIMD instructions.
\begin{enumerate}
\item
%\leoinline{I can see how this description made a referee unhappy. Here is my attempt to simplify the presentation:}
%\danielinline{Great work. I have re-ordered the the explanation somewhat so that it reads more like a continuous story. I have also removed the pronoun 'we' as we were criticized for its use in this context. (Who is 'we'?)}
%% We is editorial. I like it despite Mart Twain's joke. However, I agree that they is more appropriate here
Schlegel et al.~\cite{Schlegel2011} use a specialized data structure,
where integers are partitioned into sub-arrays of integers having the same 16 most significant bits.
To represent sub-array values,
one needs
16~bits for each sub-array element plus
16~bits to represent the most significant bytes shared among all the integers.

To compute the intersection between two lists,
Schlegel et al.\ iterate over all pairs of 16-bit sub-arrays (from
 the first and the second list, respectively)
with identical 16 most significant bits.
In each iteration, they compute the intersection over two 16-bit arrays.
To this end, each sub-array is logically divided into blocks each containing eight 16-bit integers.
For simplicity, we assume that the number of integers is a multiple of eight.
In practice, lists may contain a number of integers that is not divisible by eight.
However, one can terminate the computation of the intersection with the few remaining integers
and process these integers using a more traditional intersection algorithm.

The intersection algorithm can be seen as a generalization of the
textbook intersection algorithm (akin to the merge sort algorithm).
There are two block pointers (one for each sub-array) originally pointing to the first blocks.
Using SIMD instructions,
it is possible to carry out an all-against-all comparison
of 16-bit integers between these two blocks and extract matching integers.
Schlegel et al.\ exploit the fact that
such a comparison can quickly be done
using the SSE~4.1 string-comparison instruction \texttt{pcmpestrm}.
(See \S~\ref{sec:simdinstructions}.)
Recall that this instruction takes strings of
up to eight 16-bit integers and returns an 8-bit mask indicating which 16-bit integers are present in both strings.
The number of 1s in the resulting mask (computed quickly with the \texttt{popcnt} instruction) indicates the number of matching integers. Schlegel et al.\ apply this instruction to  two blocks of eight integers (one from the first list and another from the second list).
To obtain matching integers, they use the shuffle SSSE3 instruction \texttt{pshufb}.
This instruction extracts and juxtaposes matching integers so that they follow each other without gaps.
This requires a lookup table to convert 8-bit masks produced by \texttt{pcmpestrm} (and extracted using \texttt{pextrd})
to the shuffle mask for \texttt{pshufb}.
The lookup table contains 256 128-bit shuffle masks.
The matching integers are written out and the output pointer is advanced by the number of matching integers.

Afterwards, the highest values of the two blocks are compared.
If the highest value in the first block is smaller than the highest value of the second block,
the first block pointer is advanced. If the second block contains a smaller highest value, its pointer
is advanced in a similar fashion. If the highest values are equal, the block pointers
of both lists are advanced.
When the blocks in one of the lists are exhausted, the intersection has been computed.

We slightly improved  Schlegel et al.'s implementation by replacing the \texttt{pcmpestrm} instruction with the similar \texttt{pcmpistrm} instruction. The latter is faster. Its only downside is that it assumes that the two strings of 16-bit characters are either null terminated or have a length of 8 characters. However, because the integers are sorted (even in 16-bit packed blocks), we can easily check for the presence of a zero 16-bit integer as a special case.

Schlegel et al.\ only validated their results on the intersection of
arrays that have identical lengths.
In contrast, we  work on arrays of 32-bit integers having differing lengths.

\item Katsov~\cite{Katsov2012} proposed an approach similar
to Schlegel et al.\ except that it works on arrays of 32-bit integers. We process the input arrays by considering blocks of four 32-bit integers. Given two blocks $(a,b,c,d)$ and $(a',b',c',d')$, we execute four comparisons:
\begin{itemize}
\item  $(a,b,c,d)=(a',b',c',d')$,
\item  $(a,b,c,d)=(b',c',d',a')$,
\item  $(a,b,c,d)=(c',d',a',b')$,
\item  $(a,b,c,d)=(d',a',b',c')$.
\end{itemize}
%\daniel{Here I rewrote this next sentence. The detailed explanation has become slightly redundant given our new section so I trimmed it down a bit.}
%looks great
For each comparison, the \texttt{pcmpeqd} instruction  generates four integers with value 0xFFFFFFFF or 0 depending on whether the corresponding pairs of integers are equal. %(e.g., if $a=a'$, then the first component returned by \texttt{pcmpeqd} has value 0xFFFFFFFF).
 By computing the bitwise OR of the 4~resulting masks, we know whether each value in $(a,b,c,d)$ matches one value in $(a',b',c',d')$. We can extract the corresponding 4-bit mask with the \texttt{movmskps} instruction.
The rest is similar to Schlegel et al.'s algorithm.
%Ilya Katsov
%http://highlyscalable.wordpress.com/2012/06/05/fast-intersection-sorted-lists-sse/

%Ilya Katsov
%http://highlyscalable.wordpress.com/2012/06/05/fast-intersection-sorted-lists-sse/

\end{enumerate}

%\leoinline{Great reference. Do we want to mention tree-based approaches, but state that they are not exactly applicable as we work over arrays?}

%\danielinline{Right now, we cite  K-ary search, sort, \ldots We could add more but I am not sure it helps. As we just wrote, this research direction goes back to the 1960s. There has been lots and lots of work. As far as we know, what we propose is novel. I guess that it is the point we make, and I think we make it well enough. Ultimately, someone could always say that he does not believe that we do is novel, no matter how many references we include.}
%ok
There has been
extensive work to accelerate sort or search algorithms by using SIMD instructions~\cite{Balkesen201385,Schlegel:2009:KSM:1565694.1565705}, among others. Nevertheless, we are not aware of other algorithms practically relevant to the intersection of sorted lists of integers using SIMD instructions.
%\daniel{Referee made vague claim that our algorithms were not novel. He even points out to earlier work by Schlegel, but even Schlegel does not cite \cite{Schlegel:2009:KSM:1565694.1565705} as relevant. The SIMD algorithms they use for sorting or for heaps are quite different. It is not really the same problem.}
%Yes, agree.

\subsection{Our intersection algorithms}

Both the SvS and textbook algorithms involve checking
whether an element from one list is in the other list.
When one list is substantially shorter than the other,
we typically compare a single integer from
the shorter list against several adjacent integers from the longer list.
Such comparisons can be efficiently done using SIMD operations
(see sample code in Appendix~\ref{appendix:code})
and this is the underlying idea of our  intersection algorithms. Unlike the algorithms by Schlegel et al.\ and Katsov we do not need to process a mask to determine which integers intersect since we check only one integer at a time from the short list.

When implementing the SvS algorithm over uncompressed integers,
it is convenient to be able to write out the result of the intersection directly in one of the two lists. This removes the need to allocate
additional memory to save the result of the intersection.
All our  intersection algorithms are designed to have the
\emph{output-to-input} property:
it is always safe to write the result in the shorter of the two lists,
because the algorithms never overwrite
unprocessed data. The algorithms by Schlegel et al.\ and Katsov do not have this property.

Our simplest SIMD intersection algorithm is \textsc{V1} (see Algorithm~\ref{algo:v1}): \textsc{V1} stands for ``first vectorized intersection algorithm''. It is essentially equivalent to a simple textbook intersection algorithm (\textsc{scalar}) except that we access the integers of the long lists in blocks of $\Tau$~integers. We advance in the short list one integer at a time. We then advance in the long list until we find a block of $\Tau$~integers such that the last one is at least as large as the integer in the short list.
We compare the  block of $\Tau$~different integers in the long list with the integer in the short list using SIMD instructions. If there is a match, that is, if one of the $\Tau$~integers is equal to the integer from the short list,
we append that integer to the intersection list.

\begin{figure*}[t]
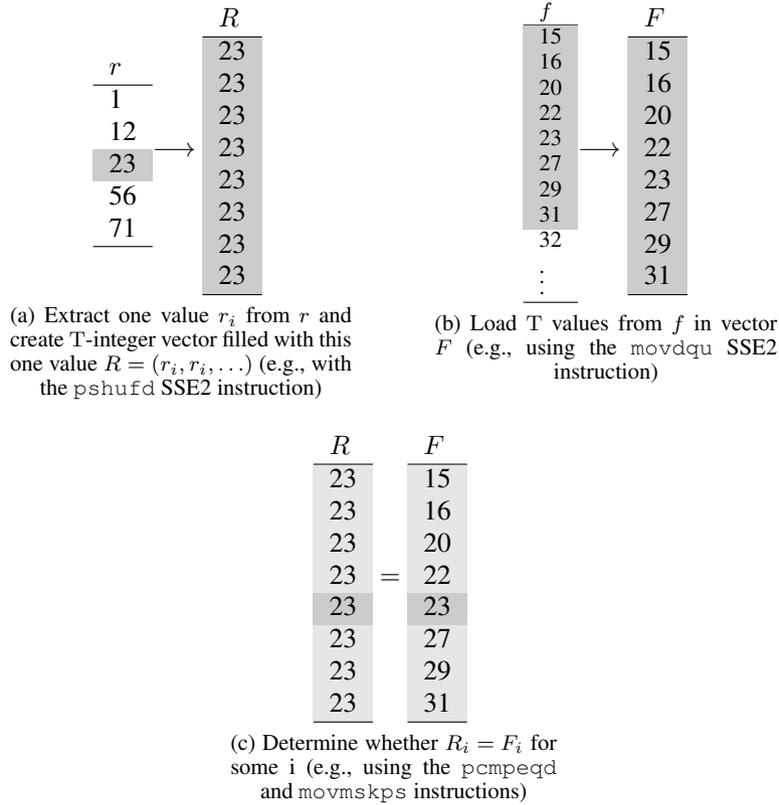

\centering
\subfloat[Extract one value $r_i$ from $r$ and create $\Tau{}$-integer vector filled with this one value
$R=(r_i,r_i,\ldots)$
(e.g., with the \texttt{pshufd} SSE2 instruction)]{%
\hspace*{1cm}
\begin{tabular}{l}
               $r$  \\\hline
               1                    \\
               12                    \\
               \cellcolor[gray]{0.8}23                   \\
               56                    \\
               71                    \\
\hline
\end{tabular}$\longrightarrow$
\begin{tabular}{l}
               $R$  \\\hline
               \cellcolor[gray]{0.8}23                    \\
               \cellcolor[gray]{0.8}23                    \\
               \cellcolor[gray]{0.8}23                   \\
               \cellcolor[gray]{0.8}23                    \\
               \cellcolor[gray]{0.8}23                    \\
               \cellcolor[gray]{0.8}23                    \\
               \cellcolor[gray]{0.8}23                    \\               \cellcolor[gray]{0.8}23                    \\
               \hline
\end{tabular}
\hspace*{1cm}
}\hspace{1cm}
\subfloat[Load $\Tau{}$ values from $f$ in vector $F$ (e.g., using the \texttt{movdqu} SSE2 instruction)]{%
\hspace*{1cm}
\begin{scriptsize}
\begin{tabular}{l}
               $f$  \\\hline
               \cellcolor[gray]{0.8}15                          \\
               \cellcolor[gray]{0.8}16     \\
               \cellcolor[gray]{0.8}20                          \\
               \cellcolor[gray]{0.8}22                           \\
               \cellcolor[gray]{0.8}23                           \\
               \cellcolor[gray]{0.8}27                           \\
               \cellcolor[gray]{0.8}29                           \\
               \cellcolor[gray]{0.8}31                           \\
               32\\
               $\vdots$                    \\
\hline
\end{tabular}\end{scriptsize}$\longrightarrow$
\begin{tabular}{l}
               $F$  \\\hline
               \cellcolor[gray]{0.8}15                          \\
               \cellcolor[gray]{0.8}16     \\
               \cellcolor[gray]{0.8}20                          \\
               \cellcolor[gray]{0.8}22                           \\
               \cellcolor[gray]{0.8}23                           \\
               \cellcolor[gray]{0.8}27                           \\
               \cellcolor[gray]{0.8}29                           \\
               \cellcolor[gray]{0.8}31                           \\
\hline
\end{tabular}
\hspace*{1cm}
}\hspace{1cm}
\subfloat[Determine whether $R_i=F_i$ for some i (e.g., using the \texttt{pcmpeqd} and \texttt{movmskps}  instructions) ]{%
\hspace{1cm}
\begin{tabular}{l}
               $R$  \\\hline
               \cellcolor[gray]{0.9}23                    \\
               \cellcolor[gray]{0.9}23                    \\
               \cellcolor[gray]{0.9}23                   \\
               \cellcolor[gray]{0.9}23                    \\
               \cellcolor[gray]{0.8}23                    \\
               \cellcolor[gray]{0.9}23                    \\
               \cellcolor[gray]{0.9}23                    \\               \cellcolor[gray]{0.9}23                    \\
               \hline
\end{tabular}
$=$
\begin{tabular}{l}
               $F$  \\\hline
               \cellcolor[gray]{0.9}15                          \\
               \cellcolor[gray]{0.9}16     \\
               \cellcolor[gray]{0.9}20                          \\
               \cellcolor[gray]{0.9}22                           \\
               \cellcolor[gray]{0.8}23                           \\
               \cellcolor[gray]{0.9}27                           \\
               \cellcolor[gray]{0.9}29                           \\
               \cellcolor[gray]{0.9}31                           \\
\hline
\end{tabular}
\hspace{1cm}
}
\caption{\label{fig:algov1}Brief illustration of Algorithm \textsc{V1} with $\Tau=8$ and $r_i=23$}
\end{figure*}

For example, consider Fig.~\ref{fig:algov1}. In the short list $r$, we want to know whether the integer 23 is present in the longer list. We start by creating a vector containing $T=8$ copies of the integer: $R=(23, 23, \ldots, 23)$. From the long list, we load 8~consecutive integers such that the last integer in the list is at least as large as 23. In this case, we load $F=(15,16,\ldots, 29,31)$. Then we compare the vectors $R$ and $F$ using as little as two SIMD operations. In our example, there is a match. There is no need to do much further work: we have determined that the integer $r$ belongs to the intersection.

%%%%%%%%%%%%%%
% basic algorithm for when the two arrays have similar lengths
%%%%%%%%%%%%%%%
\begin{algorithm}
\caption{\label{algo:v1}The \textsc{V1} intersection algorithm}
\centering
\begin{algorithmic}[1]
\REQUIRE SIMD architecture with $\Tau$-integer vectors
\STATE \textbf{input}: two  sorted non-empty arrays of integers $r, f$
\STATE \textbf{assume:} $\mathrm{length}(f)$ is divisible by $\Tau$
\STATE $x$ initially empty dynamic array (our answer)
\STATE $j \leftarrow 1$
\FOR {$i \in \{1, 2, \ldots, \mathrm{length}(r)\}$}
\STATE $R \leftarrow (r_i, r_i, \ldots, r_i)$
\WHILE{$f_{j- 1 +  \Tau }<  r_i$ }
\STATE $j \leftarrow j + \Tau$
\IF {$j> \mathrm{length}(f)$ }
\RETURN  $x$
\ENDIF
 \ENDWHILE
\STATE $F \leftarrow ( f_{j}, f_{j+1}, \ldots, f_{j-1+\Tau})$
\IF {$R_i = F_i$ for some $i\in \{1,2, \ldots, \Tau\}$ }
\STATE append $r_i$ to $x$
\ENDIF
\ENDFOR
\RETURN $x$
\end{algorithmic}
\end{algorithm}

We found it best to use the \textsc{V1} algorithm with $\Tau=8$.
Because SSE2 only offers 4-integer vectors, we simulate an 8-integer vector by using two 4-integer vectors (see Appendix~\ref{appendix:code}).
To compare two vectors for possible equality between any two components,
we use the \texttt{pcmpeqd} instruction:
%\daniel{This explanation of how pcmpeqd works is slightly redundant, but I did not edit it.}
 given 4~pairs of 32-bit integers, it generates 4~integers: \texttt{0xFFFFFFFF} when the two integers are equal and \texttt{0x00000000} otherwise. We can combine the result of two such tests
%this is fine, they sure will forget how this instruction works when they reach this point
with a bitwise OR (\texttt{por}).
To test whether there is a match of any of the integers, we use the
\texttt{movmskps} instruction.

However, the computational complexity of the \textsc{V1} algorithm is still $O(m+n/\Tau)$
where $\Tau$ is limited to small constants by the CPU architecture. Hence, a simple galloping intersection can be faster than V1 when $n\gg m$ as the galloping algorithm has complexity $O(m \log n)$.
% it was two layers ON branching. Not 100% sure if ON branching is incorrect or not

%\daniel{V3 is better explained, though it could maybe be smoother.}
%\leo{It's good, but I added a note that it's a binary search}
% Daniel : OK.
We can optimize further and add two layers of branching (henceforth \textsc{V3}, see  Algorithm~\ref{algo:v3}).
%%%
That is, like with \textsc{V1}, we first locate a  block of $4T$~integers ($f_j, \ldots, f_{j-1+4T}$) in the larger array $f$ where a match could be found for the current value $r_i$ selected from the small array: we increment $j$ by steps of $4T$ until the last value from the current block is at least as large as the value from the small array,  $f_{j-1+4T}\geq r_i$.
However, instead  of directly comparing $4T$~pairs of integers as the analog of Algorithm~\textsc{V1} would do, we use the binary search algorithm to find the one block of $T$~integers within the larger block of $4T$~integers where a match is possible.
Technically, we compare the current value from the small array $r_i$  with the value in the middle of the block of $4T$ (integers $f_{j-1+2T}$) and then again with either the value in the middle of the first half ($f_{j-1+T}$) or in the middle of the second half ($f_{j-1+3T}$).
Thus, some comparisons can be skipped. We first tried a version, called \textsc{V2}, that added only one layer of branching to \textsc{V1} but it did not prove useful in practice. %\daniel{Better explaining our names.}

However, when $n$ is large, galloping is superior to both  \textsc{V1} and  \textsc{V3}. Thus, we also created a SIMD-based galloping (henceforth \textsc{SIMD Galloping}, see Algorithm~\ref{algo:simdgalloping}).
 It uses the same ideas as galloping search, except that we exploit the fact that SIMD instructions can compare $T$~pairs of integers at once. \textsc{SIMD Galloping} has similar complexity in terms of $m$ and $n$ as  scalar galloping ($O(\frac{m}{\Tau} \log n)=O(m \log n)$) so we expect good scalability.

%\leo{It is not exactly the same as we add a factor $\frac{1}{\Tau}$}
%\daniel{Granted. I replaced "the same" by "similar".}
%ok

\subsubsection{SIMD Hybrid Algorithm}
\label{sec:hybrid}

In practice, we find that our \textsc{SIMD Galloping} is always faster than a non-SIMD galloping implementation.
Nevertheless, to fully exploit the speed of SIMD instructions, we find it desirable to still use \textsc{V1} and  \textsc{V3} when they are faster. Thus, when processing 32-bit integers, we use
a combination of \textsc{V1}, \textsc{V3}, and SIMD galloping
where a choice of the intersection algorithm is defined by the following heuristic:

\begin{itemize}
\item When $\mathrm{length}(r) \leq \mathrm{length}(f)<50 \times \mathrm{length}(r)$,
we use  the \textsc{V1} algorithm with 8-integer vectors ($\Tau=8$, see Algorithm~\ref{algo:v1}).
\item When $50 \times \mathrm{length}(r) \leq \mathrm{length}(f) < 1000 \times \mathrm{length}(r)$, we use  the \textsc{V3} algorithm with 32-integer vectors ($\Tau=32$, see Algorithm~\ref{algo:v3})

\item When $1000 \times \mathrm{length}(r) \leq \mathrm{length}(f)$, we use  the \textsc{SIMD Galloping} algorithm with 32-integer vectors ($\Tau=32$, see Algorithm~\ref{algo:simdgalloping}).
\end{itemize}
Though we are most interested in 32-bit integers, \textsc{V1}, \textsc{V3} and \textsc{SIMD Galloping}  can also be used with arrays of 8-bit, 16-bit and 64-bit integers. Different heuristics would be needed in such cases.
%%\leo{As well as in the case of collections substantially different from ClueWeb09 and Gov2?}\daniel{We have not yet discussed ClueWeb09 and Gov2. I recommend we keep things simple. Yes, we could need different heuristics for different application domains, but if we start going there, referees will ask to explain more and more. It is a trap!}.
%% Ok, but in the general case, of course, one would want to tune parameters for better performance. It is actually good, we got this parameters from synthetic data.
%% Thus, nobody will claim we overfit to a specific set. However, our results might be a bit suboptimal because of this. One can probably obtain better performance in a cross-validated experiment. Though this would be a bit of pain.

%We could also come up with many variations similar to Algorithm \textsc{V3}: some using more branches, some less, some using longer vectors some using short ones.
%Though we expect that it is possible to gain further speed by either tuning our chosen algorithms or adding new variations, we may also get diminishing returns with increasing complexity. We leave such investigations to future work.
%\daniel{Algorithm 3 jumps with a stride of 4T to skip data. There is a rich design space by this paradigm: for example, you can jump 2T with one less branch or 8T with one more branch. It is not justified why jumping 4T is optimal.}

\subsubsection{Possible refinements}

Our SIMD Intersection algorithms assume that the longer array has length divisible by either $\Tau$ or $4 \Tau$. In practice, when this is not the case, we  complete the computation of the intersection with the \textsc{scalar} algorithm. We expect that the contribution of this step to the total running time is small. As a possibly faster alternative, we could pad the longer array to a desired length using a special integer value that is guaranteed not to be in the shorter array.  We could also fallback on other SIMD algorithms to complete the computation of the intersection.
%\leoinline{I don't get the last statement. Do we fallback each time when the length is not divisible?  Note that this sentences is also copied to the response document.} \danielinline{I tried to make it more precise by adding ``to complete the computation of the intersection''. The idea of the referee is that we could fall back on V1 after using SIMD Galloping and having some leftovers. I am trying not to say too much because if we expand too much, we risk building up the expectation that we have to benchmark these techniques.}
%% ok, I completely missed that this is a fallback for the remaining parts of the lists.

We could introduce other refinements. For example, our implementation of  \textsc{SIMD Galloping} relies on a scalar binary search like
 the conventional galloping. However, recent Intel processors have introduced gather instructions (\texttt{vpgatherdd}) that can retrieve at once several integers from various locations~\cite{Polychroniou:2014:VBF:2619228.2619234}. Such instructions could accelerate binary search. We do not consider such possibilities further.

%%%%%%%%%%%%%%
% fancier algorithm for when the two arrays have moderately different lengths
%%%%%%%%%%%%%%%
\begin{algorithm}[bth]
\caption{\label{algo:v3}The \textsc{V3} intersection algorithm}
\centering
\begin{algorithmic}[1]
\REQUIRE SIMD architecture with $\Tau$-integer vectors
\STATE \textbf{input}: two sorted non-empty arrays of integers $r, f$
\STATE \textbf{assume:} $\mathrm{length}(f)$ is divisible by $4 \Tau$
\STATE $x$ initially empty dynamic array (our answer)
\STATE $j \leftarrow 1$
\FOR {$i \in \{1, 2, \ldots, \mathrm{length}(r)\}$}
\STATE $R \leftarrow (r_i, r_i, \ldots, r_i)$ \\
 \WHILE{$f_{j - 1+ 4 \Tau }< r_i$ }
\STATE $j \leftarrow j + 4 \Tau$
\IF {$j> \mathrm{length}(f)$ } \RETURN $x$\ENDIF
 \ENDWHILE
\IF{$f_{j-1+2\Tau }\geq r_i$}
\IF{$f_{j-1+\Tau }\geq r_i$}
\STATE $F \leftarrow ( f_{j}, f_{j+1}, \ldots, f_{j-1+\Tau})$
\ELSE
\STATE $F \leftarrow ( f_{j+\Tau}, f_{j+2+\Tau}, \ldots, f_{j-1+2\Tau})$
\ENDIF
\ELSE
\IF{$f_{j-1+3\Tau }\geq r_i$}
\STATE $F \leftarrow ( f_{j+2\Tau}, f_{j+2+2\Tau}, \ldots, f_{j-1+3\Tau})$
\ELSE
\STATE $F \leftarrow ( f_{j+3\Tau}, f_{j+2+3\Tau}, \ldots, f_{j-1+4\Tau})$
\ENDIF
\ENDIF
\IF {$R_i = F_i$ for some $i\in \{1,2, \ldots, \Tau\}$ }
\STATE append $r_i$ to $x$
\ENDIF
\ENDFOR
\RETURN  $x$
\end{algorithmic}
\end{algorithm}

%%%%%%%%%%%%%%
% fancier algorithm for when the two arrays have moderately different lengths
%%%%%%%%%%%%%%%
\begin{algorithm}
\caption{\label{algo:simdgalloping}The \textsc{SIMD~Galloping} algorithm}
\centering
\begin{algorithmic}[1]
\REQUIRE SIMD architecture with $\Tau$-integer vectors
\STATE \textbf{input}: two non-empty sorted arrays of integers $r, f$
\STATE \textbf{assume:} $\mathrm{length}(f)$ is divisible by $\Tau$
\STATE $x$ initially empty dynamic array (our answer)
\STATE $j \leftarrow 1$
\FOR {$i \in \{1, 2, \ldots, \mathrm{length}(r)\}$}
\STATE $R \leftarrow (r_i, r_i, \ldots, r_i)$
\STATE \textbf{find} by sequential search the smallest $\delta$ such that $f_{j+\delta-1+ \Tau  }\geq r_i$ for $\delta = 0, 1 \Tau, 2\Tau, 4\Tau, \ldots, \mathrm{length}(f)-1+\Tau$,  \textbf{if} none \textbf{return}~$x$
\STATE \textbf{find} by binary search the  smallest $\delta_{\mathrm{min}}$ in $[\lfloor \delta/2 \rfloor,\delta]$ divisible by $\Tau$ such that $f_{j+\delta_{\mathrm{min}}-1+ \Tau  }\geq r_i$
\STATE $j \leftarrow j + \delta_{\mathrm{min}}$
\STATE $F \leftarrow ( f_{j}, f_{j+1}, \ldots, f_{j-1+\Tau})$
\IF {$R_i = F_i$ for some $i\in \{1,2, \ldots, \Tau\}$ }
\STATE append $r_i$ to $x$
\ENDIF
\ENDFOR
\RETURN  $x$
\end{algorithmic}
\end{algorithm}

\section{Experimental results}
We  assess experimentally the unpacking speed
with \emph{integrated} differential coding (\S~\ref{sec:expbitunpacking}), the decompression speed of the
corresponding  schemes (\S~\ref{sec:expdecoding}), the benefits of our SIMD-based intersection schemes (\S~\ref{sec:expintersections}), and, SIMD-accelerated bitmap-list hybrids (\S~\ref{sec:exphyb}) with realistic query logs and inverted indexes.

\subsection{Software}\label{SectSoftware}
All our software is freely available online\footnote{\anonymizeurl{https://github.com/lemire/SIMDCompressionAndIntersection}.
} under the Apache Software License~2.0.
The code is written in C++ using the C++11 standard.
Our code builds using several compilers such as
\texttt{clang++}~3.2 and Intel \texttt{icpc}~13. However, we
use \texttt{GNU~GCC}~4.7 on a Linux PC for our tests. All code was compiled using the \texttt{-O3} flag.
 % By design, all our software is single-threaded.
We  implemented scalar schemes (\textsc{FastPFOR} and \textsc{varint}) without using SIMD instructions and with the scalar equivalent of \textsc{D1} differential coding.
%\leo{It is not only for clarity, I guess, it is also to have fair comparison of SIMD vs scalar implementation. Do we want to elaborate on this a bit?}

%\daniel{I have removed the "for clarity" but I do not think we should elaborate. It is a reasonable choice.}
%ok

\subsection{Hardware}

We ran all our experiments on an
Intel Xeon CPU (E5-1620, \emph{Sandy Bridge}) running at \SI{3.60}{\GHz}.
%This CPU has 4~cores, but we expect that a single core is used during
%our benchmarks.
This CPU also has \SI{10}{\MB} of L3 cache as well as
\SI{32}{\kB} and \SI{256}{\kB} of L1 and L2 data cache per core.
We have \SI{32}{\GB} of RAM (DDR3-1600) running quad-channel.
We estimate that we can read from RAM at a speed of 4~billion 32-bit integers
per second and from L3~cache at a speed of 8~billion 32-bit integers per second. All data is stored in RAM so that disk performance is irrelevant.

\subsection{Real data}\label{SectRealData}
To fully assess our results, we need realistic data sets. For
this purpose, we use posting lists extracted from the
TREC collections ClueWeb09 (Category B) and GOV2.
Collections GOV2 and ClueWeb09 (Category B) contain roughly 25~and 50~million HTML documents, respectively.
These documents were crawled from the web. In the case of GOV2 almost all pages were collected
from websites in either the \texttt{.gov} or \texttt{.us} domains,
but the ClueWeb09 crawl was not limited to any specific domain.

ClueWeb09 is a partially sorted collection: on average it has runs of 3.7~thousand
documents sorted by URLs. In GOV2, the average length of the sorted run is almost one.
Thus, we use two variants of GOV2: the original and a sorted one, where documents
are sorted by their URLs.
We did not consider other sorting strategies such as
by number of terms in the document (terms-in-document) or by document size~\cite{kane2014,Tonellotto:2011:EDD:2009916.2010108}.

We first indexed collections using Lucene (version~4.6.0):
the words were stopped using
the default Lucene settings, but not stemmed.
Then, we extracted postings corresponding to one million most frequent terms.
In both GOV2 and ClueWeb09, the excluded terms represent \SI{4}{\percent} of the postings. Morever, the excluded posting lists had averages of 3.2~postings for GOV2 and 4.5~postings  for ClueWeb09.
%
%ClueWeb09: Term qty: 150410798 Doc qty: 50312694 postings qty: 16317942611
%Gov2:           Term qty: 74783244   Doc qty: 25205179 postings qty: 6218290684
%
%Cut:
%
%ClueWeb09: Term qty: 1000000  postings qty: 15641166521
%Gov2:            Term qty: 1107205 of postings: 5979715441
%
The extracting software as well as the processed data is freely available online.\footnote{\anonymizeurl{https://github.com/searchivarius/IndexTextCollect} and \anonymizeurl{http://lemire.me/data/integercompression2014.html}.}
Uncompressed, extracted posting lists from GOV2 and ClueWeb09 use 23\,GB and 59\,GB, respectively.
They include only document  identifiers.

Our corpora represent realistic sets of documents obtained by splitting a large collection
 so that each part fits into the memory of a single server.
In comparison, other researchers use collections of similar or smaller sizes.
Culpepper and Moffat~\cite{Culpepper:2010:ESI:1877766.1877767}, Ao et al.~\cite{Ao:2011:EPL:2002974.2002975}, Barbay et al.~\cite{Barbay:2010:EIS:1498698.1564507}, Ding et al.~\cite{Ding:2009:UGP:1526709.1526766} and  Vigna~\cite{Vigna2013} used TREC GOV2 (25M documents),
Ding and K\"onig~\cite{Ding:2011:FSI:1938545.1938550} indexed Wikipedia (14M documents) while Transier and Sanders~\cite{Transier:2010:EBA:1877766.1877768} used WT2g (250k documents).

\subsubsection{Query logs}
\label{sec:querylogs}

%no more AOL
In addition, we used the TREC million-query track~(1MQ)~logs (60~thousand queries from years 2007--2009).
We randomly picked 20~thousand queries containing at least two indexed terms.
These queries were converted into sequences of posting  identifiers for GOV2 and ClueWeb09.
Recall that postings beyond one million most frequent terms were eliminated.
Therefore, we had to exclude \SI{4.6}{\percent} of queries in the case of ClueWeb09
and \SI{8.4}{\percent} of queries in the case of GOV2 (which could not be converted
to posting identifiers).
We believe that excluding  a small fraction of the queries slightly biased up processing times,
but the overall effect was not substantial.

Table~\ref{table:querylogstats} gives intersection statistics for the GOV2 and Clueweb09 collections.
We give the percentage of all queries having a given number of terms,
the average
size of the intersection, along with the average size of the smallest posting list (term 1), the average size of the
second smallest (term 2) and so on.

To insure that all our indexes fit in RAM, we index only the terms used by the given query log. When reporting compression results (in bits/int), we only consider this in-memory index.

\begin{table}[tbh]
\caption{Statistics about the TREC Million-Query log on two corpora.
We give the percentage of all queries having a given number of terms, the average size of the intersection along with the average size of the smallest posting list (term 1), the average size of the second smallest (term 2) and so on. All sizes are in thousands.
There are~50M documents in total for Clueweb09 and half as many for Gov2.\label{table:querylogstats}
}\centering
\subfloat[Clueweb09  (matching documents in thousands)]{%
\begin{tabular}{c|c|c|l}\hline
\# & \% & inter. & term 1, term 2, \ldots \\\hline
2	&	19.8	&	93&	\num{380},	\num{2600}\\
3	&	32.5	&	29&	\num{400},	\num{1500},	\num{5100}\\
4	&	26.3	&	17&	\num{480},	\num{1400},	\num{3200},	\num{8100}\\
5	&	13.2	&	12&	\num{420},	\num{1200},	\num{2600},	\num{4800},	\num{10000}\\
6	&	4.9	&	4	&	\num{350},	\num{1000},	\num{2100},	\num{3700},	\num{6500},	\num{13000}\\
7	&	1.7	&	5	&	\num{390},	\num{1100},	\num{2100},	\num{3400},	\num{5200},	\num{7300},	\num{13000}\\
\end{tabular}
}\\
\subfloat[Gov2  (matching documents in thousands)]{%
\begin{tabular}{c|c|c|l}\hline
\# & \% & inter. & term 1, term 2 , \ldots \\\hline
2	&	19.6	&	49 &	\num{160},	\num{1100}\\
3	&	32.3	&	22 &	\num{180},	\num{710},	\num{2400}\\
4	&	26.4	&	11 &	\num{210},	\num{620},	\num{1500},	\num{3700}\\
5	&	13.4	&	7&	\num{170},	\num{520},	\num{1100},	\num{2200},	\num{4400}\\
6	&	5.0	&	3	&	\num{140},	\num{420},	\num{850},	\num{1500},	\num{2600},	\num{5100}\\
7	&	1.8	&	9	&	\num{190},	\num{440},	\num{790},	\num{1300},	\num{2200},	\num{3200},	\num{5400}\\
\end{tabular}
}
\end{table}

\subsection{Bit unpacking}\label{sec:expbitunpacking}

Bit unpacking (see \S~\ref{sec:bitpackingonly}) is the fundamental operation we use as building blocks for our compression schemes (i.e., the families \textsc{S4-BP128}, \textsc{FastPFOR}, \textsc{S4-FastPFOR}).  In our experiments,
unpacking always includes differential coding, as an integrated operation or a separate step (see \S~\ref{sec:diffcod}).

%\leo{I suspect that $ \leq x_{i+4}-x_i$ is confusing. I better say that gaps were simply from 0 to $2^b$}
%\daniel{I have simplified it with ``smaller than $2^b$''. }
% Good!
To test the speed of our various bit unpacking routines, we generated
random arrays of $4096$~32-bit integers for each bit width $b= 1,2, \ldots, 31$.
We generated gaps smaller than $2^b$ %$0 \leq x_{i+4}-x_i < 2^b$
using the C \texttt{rand} function as a pseudo-random number generator.
Because SIMD unpacking routines operate on blocks of 128~integers (see Algorithm~\ref{algo:unpacking}), a total of $4096/128=32$~calls to the same unpacking routine is required to unpack an entire array.
 For each bit width, we ran $2^{14}$~sequences of  packing and unpacking and report only the average.

\begin{figure}
\centering
\includegraphics[width=0.7\columnwidth]{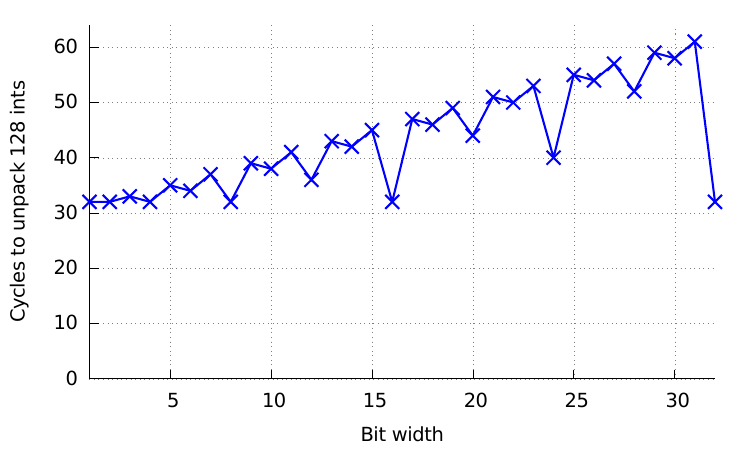}
\caption{\label{fig:iaca}Estimated reciprocal throughput of the SIMD unpacking procedures according to IACA (for 128~32-bit integers and without differential coding).}
\end{figure}

Generally, unpacking is fastest for low bit widths and for bit widths that are powers of two. We used the
Intel Architecture Code Analyzer (IACA) on our 32~SIMD unpacking routines---omitting  differential coding for simplicity.  IACA provides an optimistic estimation of the reciprocal throughput of these procedures: a value of 32~cycles means that if we repeatedly called the same routine, one full routine might complete every 32~cycles. Thus, in our case, we get an optimistic estimate of the number of cycles it takes to unpack 128~integers. Fig.~\ref{fig:iaca} provides this reciprocal throughput for all bit widths on our processor microarchitecture (\emph{Sandy Bridge}). It takes between 32 to 61~cycles to execute each 128-integer unpacking procedure, and the estimated throughput depends on the  bit width.
IACA reports that
 execution port~4 is a bottleneck for bit widths 1, 2, 3, 4, 8, 16, 32: it indicates that we are limited by register  stores in these cases. Indeed, 32~128-bit store operations are required to write 128~32-bit integers and we can only issue one store instruction per cycle: we therefore need 32~cycles.
 Otherwise, execution port~0 is a bottleneck. This port is used by shift instructions on \emph{Sandy Bridge}: these instructions are used more often when bit widths are large, except if the bit width is a power of two.

\begin{figure}[t]
\centering
\subfloat[One-pass (integrated) vs.\ two-pass (block-level)  prefix sum \label{fig:integratedvsnot}]{\centering\includegraphics[width=0.7\columnwidth]{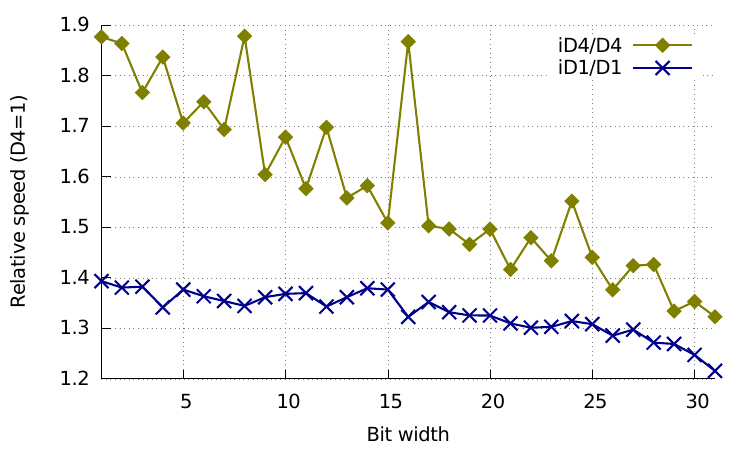}}\\
\subfloat[Several differential bit unpacking speeds\label{fig:allofthem}]{\centering\includegraphics[width=0.7\columnwidth]{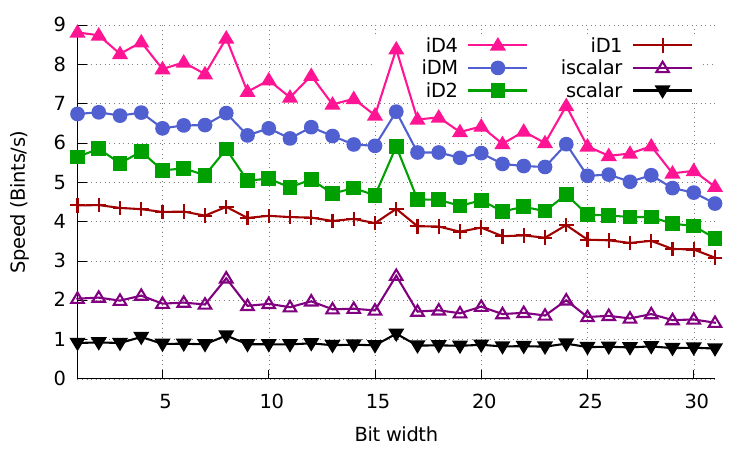}}
\caption{Unpacking speed for all bit widths ($b=1, \ldots, 31$). See Algorithm~\ref{algo:unpacking}. Speed is reported in billions of 32-bit integers per second. We use  arrays of  $4096$~integers for these tests. }
\end{figure}

In Fig.~\ref{fig:integratedvsnot}, we plot the speed ratios of the integrated bit unpacking (i.e., Algorithm~\ref{algo:unpacking} for SIMD routines),  vs.\ the regular one where differential coding is applied separately on small blocks of 32-bit integers.
To differentiate the integrated and regular differential coding we use an i prefix (i\textsc{D4} vs.\ \textsc{D4}).
For \textsc{D4},  integration improves the speed by anywhere from over
\SI{30}{\percent} to \SI{90}{\percent}.
 For \textsc{D1}, the results
are less impressive as the gains range from \SI{20}{\percent} to \SI{40}{\percent}. For  \textsc{D2} and  \textsc{DM}, the result lies in-between. Moreover, the gains due to the integration  are most important when the unpacking is fastest: for small bit widths or bit widths that are a power of two such as 8 and 16. This phenomenon is consistent with our analysis (see \S~\ref{sec:diffcod}): the benefits of integration are mostly due to the reduction by half of the number of registers stored to memory, and these stores are more of a bottleneck for these particular bit widths.

In Fig.~\ref{fig:allofthem},
we compare all of the integrated bit unpacking procedures (i\textsc{D4}, i\textsc{DM}, i\textsc{D2}, i\textsc{D1}). As our analysis predicted (see Table~\ref{table:compd}), \textsc{D4} is the fastest followed by \textsc{DM}, \textsc{D2} and \textsc{D1}. For comparison, we also include the scalar bit unpacking speed. The integrated bit unpacking (iscalar curve) is sometimes  twice as fast as regular bit unpacking (scalar curve). Even so, the fastest scalar unpacking routine has half the speed as our slowest SIMD unpacking routine (i\textsc{D1}).

\subsection{Decompression speed}\label{sec:expdecoding}

We consider several fast compression schemes: \textsc{Varint} (see \S~\ref{sec:varint}), \textsc{S4-BP128} (see \S~\ref{sec:s4bp128}),
\textsc{FastPFOR} and \textsc{S4-FastPFOR} (see \S~\ref{sec:fastpfor}).
To test the  speed of our integer compression schemes, we generated arrays using the
ClusterData distribution from  Anh and Moffat~\cite{Anh:2010:ICU:1712666.1712668}.
This distribution primarily leaves small gaps between successive
integers, punctuated by occasional larger gaps.
We generated arrays of 65536~32-bit integers in either
$[0,2^{19})$ or $[0,2^{30})$. In each case, we give the decompression speed of our schemes,  the entropy of the deltas  as well as the speed (in billions of 32-bit integers per second) of a simple copy (implemented as a call to \texttt{memcpy}) in Table~\ref{sec:decodingspeed}.
All compression schemes use differential coding.
We append a suffix (\textsc{-D4},  \textsc{-DM},  \textsc{-D2}, and  \textsc{-D1}) to indicate the type of differential coding used.
We append the \textsc{-NI} suffix (for \emph{non-integrated}) to \textsc{S4-BP128-*}
schemes where the prefix sum requires a second pass.
We get our best speeds with \textsc{S4-BP128-D4} and \textsc{S4-BP128-DM}.\footnote{Though we report a greater speed in one instance for \textsc{S4-BP128-DM} (5.5) than for \textsc{S4-BP128-D4} (5.4), the 3-digit values are  close:   5.45 and 5.44.} Given our  \SI{3.60}{\GHz} clock speed, they decompress 32-bit integers using between 0.7 and 0.8~CPU cycles per integer. In contrast, the best results reported by  Lemire and Boytsov~\cite{LemireBoytsov2013decoding} was 1.2~CPU cycles per 32-bit integer, using similar synthetic data with a CPU of the same family (\emph{Sandy Bridge}).

\textsc{S4-BP128-D1-NI}
 is  \SI{38}{\percent} faster
than  S4-FastPFOR
but  S4-FastPFOR compensates with a superior
compression
ratio (\SI{5}{\percent}--\SI{15}{\percent} better).
However, with integration, \textsc{S4-BP128-D1} increases the speed
gap to \SI{50}{\percent}--\SI{75}{\percent}.

%\daniel{One referee seemed to reject to value of our work because of the low speed of RAM\@. Let us be clear on this.}
%On our test machine, we can read from RAM at a speed of only 4\,Bint/s whereas \textsc{S4-BP128-D4} can decode integers at an even greater speed. Such a good speed is only beneficial if we only decode integers in CPU cache.

\begin{table}
\caption{\label{sec:decodingspeed}Results on ClusterData for dense ($2^{16}$ integers in $[0,2^{19})$) and
sparse ($2^{16}$ integers in $[0,2^{30})$). We report decompression speed in billions of 32-bit integers per second (Bint32/s). Shannon entropy of the deltas is given. Our Intel CPU runs at \SI{3.6}{\GHz}. The \textsc{-NI} suffix (for \emph{non-integrated}) indicates that the prefix sum requires a second pass over 128-integer blocks. The standard deviation of our timings is less than \SI{5}{\percent}.
}
\centering
\begin{tabular}{l|c@{\hskip 0.5em}c|c@{\hskip 0.5em}c}\hline
&\multicolumn{2}{c}{dense}&\multicolumn{2}{c}{sparse}\\
 & bits/int & Bint32/s& bits/int & Bint32/s \\\hline\hline
entropy & 3.9 & -- &14.7 & --\\
copy & 32.0 & 5.4 & 32.0 & 5.4 \\
\hdashline[1pt/1pt]
\textsc{S4-BP128-D4} & 6.0 & 5.4 & 16.5 & 4.4 \\
\textsc{S4-BP128-D4-NI} & 6.0 & 3.9 & 16.5 & 3.3 \\\hdashline[1pt/1pt]
\textsc{S4-BP128-DM} & 5.9 & 5.5 & 16.3 & 4.1 \\
\textsc{S4-BP128-DM-NI} & 5.9 & 3.9 & 16.3 & 3.3 \\
\hdashline[1pt/1pt]
\textsc{S4-BP128-D2} & 5.5 & 4.8 & 16.0 & 3.5 \\
\textsc{S4-BP128-D2-NI} & 5.5 & 3.7 & 16.0 & 3.2 \\
\hdashline[1pt/1pt]
\textsc{S4-BP128-D1} & 5.0 & 3.9 & 15.5 & 3.0 \\
\textsc{S4-BP128-D1-NI} & 5.0 & 3.0 & 15.5 & 2.7 \\
\hdashline[1pt/1pt]
\textsc{S4-FastPFOR-D4} & 5.8 & 3.1 & 16.1 & 2.6 \\
\textsc{S4-FastPFOR-DM} & 5.5 & 2.8 & 15.8 & 2.4 \\
\textsc{S4-FastPFOR-D2} & 5.1 & 2.7 & 15.4 & 2.4 \\
\textsc{S4-FastPFOR-D1} & 4.4 & 2.2 & 14.8 & 2.0 \\\hdashline[1pt/1pt]
\textsc{FastPFOR} & 4.4 & 1.1 & 14.8 & 1.1 \\
\hdashline[1pt/1pt]
\textsc{varint} & 8.0 & 1.2 & 17.2 & 0.3 \\
\end{tabular}
\end{table}

\subsection{Intersections}\label{sec:expintersections}

We present several intersections algorithms between (uncompressed) lists of 32-bit integers in \S~\ref{sec:fastintersection}.
To test these intersections, we again generate lists using the ClusterData distribution~\cite{Anh:2010:ICU:1712666.1712668}: all lists contain integers in $[0,2^{26})$.
 Lists are generated in pairs.
First, we set the target cardinality $n$ of the largest lists to $2^{22}$. Then we vary
the target cardinality $m$ of the smallest list from $n$ down to $n/10000$.
To ensure that intersections are not trivial, we first generate an ``intersection'' list of size $m/3$ (rounded to the nearest integer). The smallest list is built from the union of the intersection list with another list made of $2m/3$~integers.
Thus, the maximum length of this union is $m$. The longest list is made of the union of the intersection list with another list of cardinality $n-m/3$, for a total cardinality of up to $n$. The net result is that we have two sets with cardinalities $\approx m$ and $\approx n$ having an intersection at least as large as $m/3$. We compute the average of 5~intersections (using 5~pairs of lists). This choice ($m/3$) is motivated by our
experience with search engine queries (e.g., see Table~\ref{table:querylogstats}) where the intersection size is often about \SI{30}{\percent} of the smaller of two sets
%\daniel{We had ``smallest of two sets'' but referee demanded ``smaller of two sets.''}
% nitpicker detected, but who cares, easier to fix than argue
for 2-term queries.  We also present the figures for the case where the intersection is much smaller ($0.01m$): the relative results are similar. The performance of the intersection procedures is sensitive to  the
data distribution however. When we replaced ClusterData with a uniform distribution,
the intersection speed  diminished by up to a factor of 2.5: % run figures $ python compareblocks.py
value clustering  improves branch predictions
and skipping~\cite{yan2009inverted}.

\begin{figure}
\centering
\subfloat[SIMD V1 vs.\ SIMD V3 vs.\ SIMD Galloping ($m/3$ left, $0.01m$ right)\label{fig:all3simd}]{\centering\includegraphics[width=0.49\columnwidth]{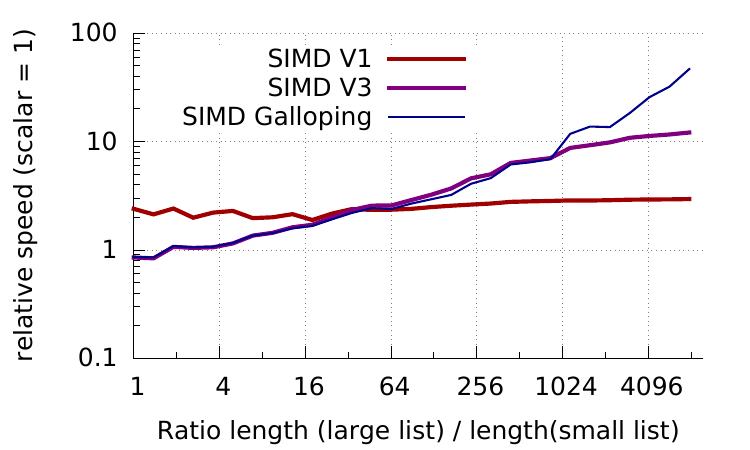}\includegraphics[width=0.49\columnwidth]{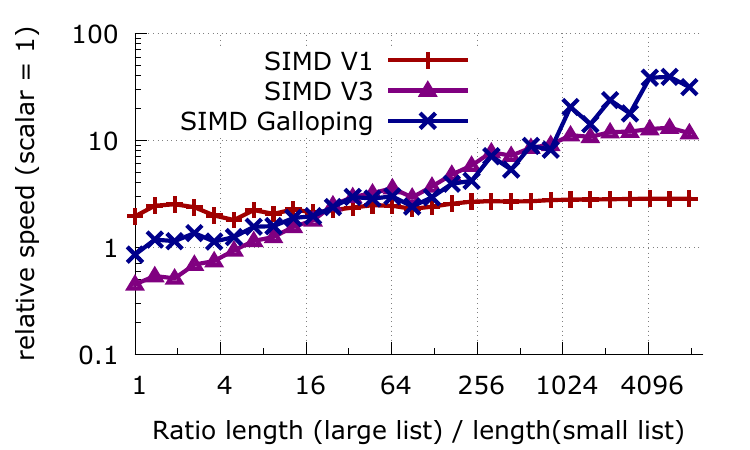}}
\\
\subfloat[SIMD  vs.\ non-SIMD galloping Intersection vs.\ Katsov ($m/3$ left, $0.01m$ right)\label{fig:simdvsgalloping}]{\centering\includegraphics[width=0.49\columnwidth]{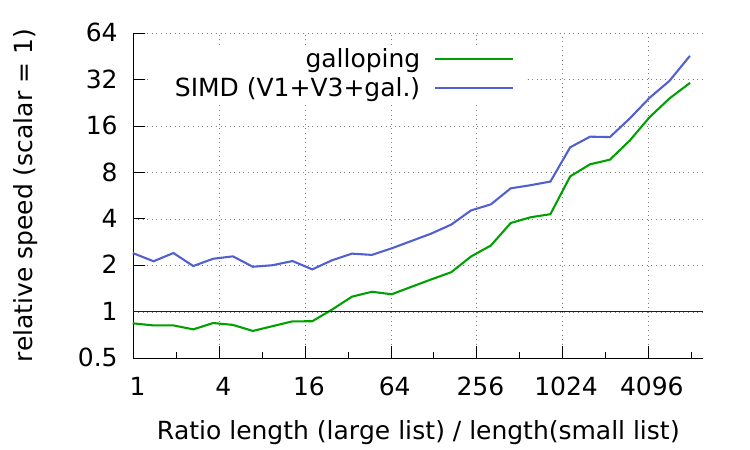}
\includegraphics[width=0.49\columnwidth]{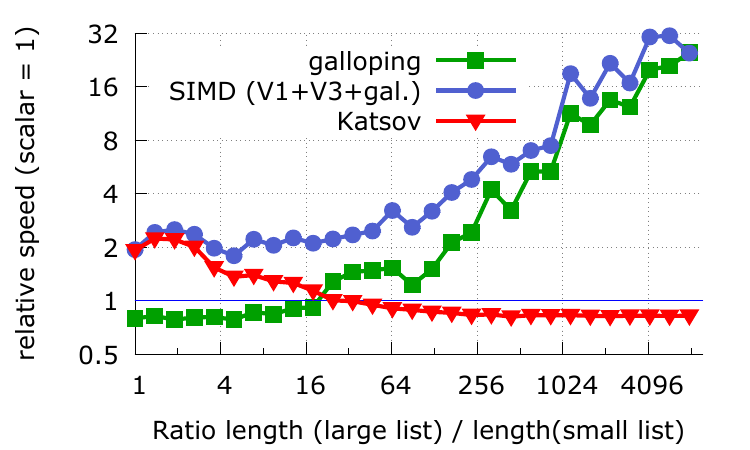}}
\\
\subfloat[16-bit case ($m/3$ left, $0.01m$ right)\label{fig:Schlegelvsus}]{\centering\includegraphics[width=0.49\columnwidth]{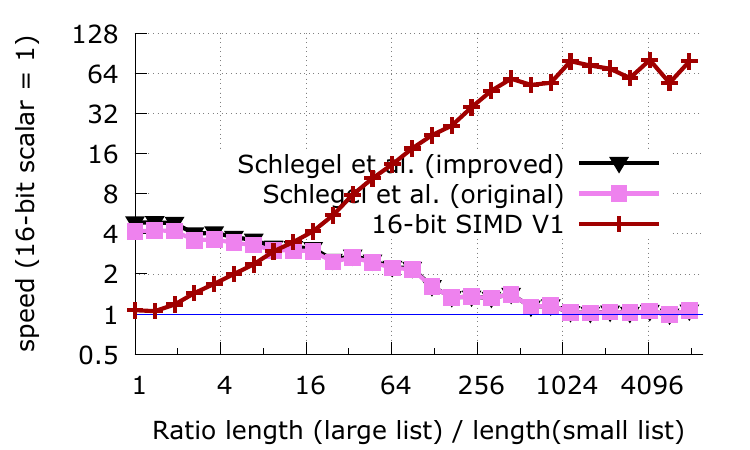}\includegraphics[width=0.49\columnwidth]{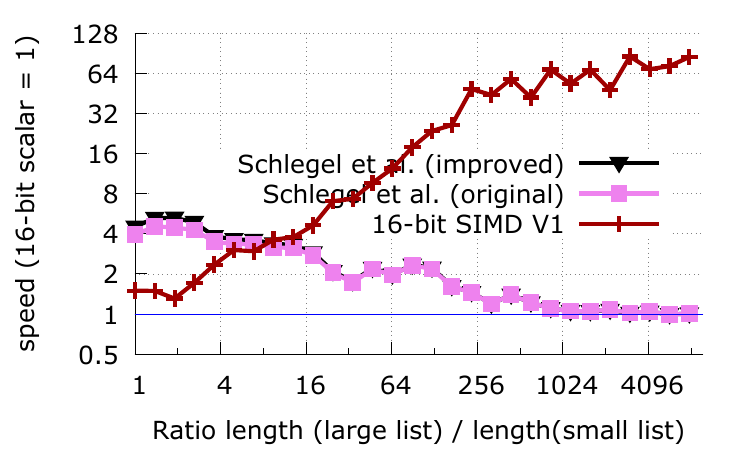}}\caption{Intersection between two lists of different cardinality as described in \S~\ref{sec:expintersections}. We provide both the case where the size of the intersection is $m/3$ and where the size of the intersection is $0.01m$.}
\end{figure}

We report all speeds relative to the basic \textsc{scalar} intersection.
All input lists are uncompressed.
%\daniel{Referee has us specify that lists are uncompressed.}
%This is, indeed, a good thing to do
In Fig.~\ref{fig:all3simd}, we compare the speed of our 3~SIMD intersection functions: \textsc{V1}, \textsc{V3}, and \textsc{SIMD Galloping}. We see that \textsc{V1} is the fastest for ratios of up to 16:1, whereas \textsc{SIMD Galloping} is the fastest for ratios larger than 1024:1. In-between, \textsc{V3} is sometimes best. This justifies our heuristic which uses \textsc{V1} for ratios up to 50:1, \textsc{V3} for ratios up to 1000:1, and \textsc{SIMD Galloping} for larger ratios. In Fig.~\ref{fig:simdvsgalloping}, we compare the two non-SIMD intersections (\textsc{scalar} and galloping) with our SIMD intersection procedure.
We see that for a wide range of cases (up to a ratio of 64:1), our SIMD intersection procedure is clearly superior to the scalar galloping, being up to twice as fast. As the ratio increases, reflecting a greater difference in list sizes,  non-SIMD galloping eventually becomes nearly as fast as our SIMD intersection.
We also compare with Katsov's algorithm
(see Fig.~\ref{fig:simdvsgalloping}).
Katsov is only competitive with  our algorithms when the lists have similar lengths.

 We also included our versions of
 Schlegel et al.'s algorithm: the  original and an improved version relying on the \texttt{pcmpistrm} instruction (see Fig.~\ref{fig:Schlegelvsus}). In these particular tests, the 32-bit integers are first transformed into Schlegel et al.'s specialized data structure, as several arrays of 16-bit integers. We also output the answer in this format.
The improved version of Schlegel et al.'s algorithm is  about \SI{15}{\percent} faster, in the best case. This supports our claim that  the \texttt{pcmpistrm} instruction is preferable to the \texttt{pcmpestrm} instruction.
We compared Schlegel et al.\ with a version of \textsc{V1} adapted to lists of 16-bit integers. Like the 32-bit  \textsc{V1}, the 16-bit \textsc{V1} compares  blocks of 256~bits and thus compares sixteen pairs of 16-bit integers at a time (i.e., $T=16$). We find that Schlegel et al.\
is better than \textsc{V1} as long as the ratio of lengths is less than 10. When one list is much longer than the other one, \textsc{V1} becomes much faster (up to $75\times$ faster in this test).
We do not use Schlegel et al.'s 16-bit format
further, but our results suggest that we could combine \textsc{V1}  and   Schlegel et al.'s algorithm for best results. That is, we could use Schlegel et al.\ for lists having similar lengths, and \textsc{V1} otherwise.

\begin{figure}
\centering
\includegraphics[width=0.7\columnwidth]{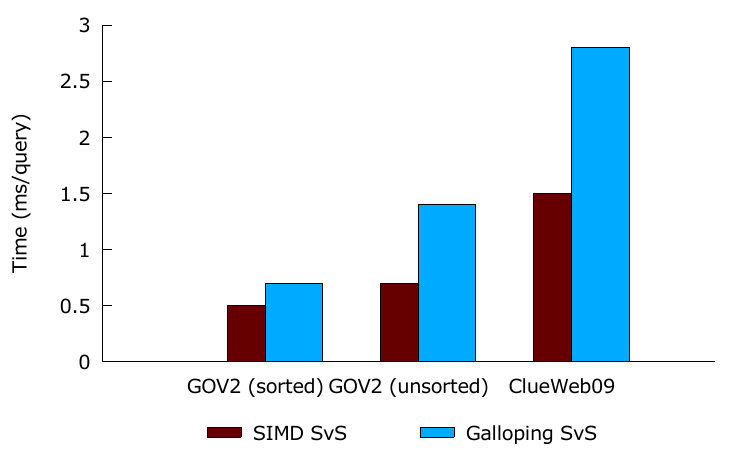}
\caption{\label{fig:comparaisoninter}Average time required to answer queries over uncompressed posting lists.}
\end{figure}

We also evaluated our algorithms using postings lists for our three collections
 using our TREC~1M Query log (see Table~\ref{table:comparaisoninter} and Fig.~\ref{fig:comparaisoninter}). For this test, posting lists are uncompressed. We see that our SIMD intersection routine is nearly twice as fast as (non-SIMD) galloping.
In the sorted version of Gov2, we expect most gaps between document identifiers to be small, with a few large gaps. In contrast,
the unsorted version of Gov2 has a more uniform distribution of
gaps.
Just as  we find that intersections are faster on the  ClusterData distributions, we find that intersections are  1.4--2$\times$ faster on
the sorted Gov2 vs.\ its unsorted counterpart.
In the next section (\S~\ref{sec:exphyb}), we show that a much better speed is possible if we use bitmaps to represent some posting lists.

\begin{table}
\caption{\label{table:comparaisoninter}Time required to answer queries in  ms/query along with the storage requirement in bits/int, using the TREC Million-Query log sample from \S~\ref{sec:querylogs}. The standard deviation of our timings is less than \SI{1}{\percent}.}
\centering
\begin{tabular}{l@{\hspace{0.2em}}c@{\hspace{0.4em}}c@{\hspace{0.6em}}c@{\hspace{0.4em}}c@{\hspace{0.6em}}c@{\hspace{0.4em}}c@{\hspace{0.1em}}}\hline
scheme &
bits/int & time &
bits/int & time &
bits/int & time  \\ \hline\hline
&
\multicolumn{2}{@{\hspace{0em}}c@{\hspace{0em}}}{\textsc{GOV2} } &
\multicolumn{2}{@{\hspace{0em}}c@{\hspace{0em}}}{\textsc{GOV2} } &
\multicolumn{2}{@{\hspace{0em}}c@{\hspace{0em}}}{\textsc{ClueWeb09} } \\
&
\multicolumn{2}{@{\hspace{0em}}c@{\hspace{0em}}}{(sorted) } &
\multicolumn{2}{@{\hspace{0em}}c@{\hspace{0em}}}{(unsorted) } &
\multicolumn{2}{@{\hspace{0em}}c@{\hspace{0em}}}{} \\
\hline
SIMD SvS&        32.0   &        0.5    &        32.0   &        0.7 & 32.0 & 1.5 \\
Galloping SvS&   32.0   &        0.7    &        32.0   &        1.4 & 32.0 & 2.8 \\
\textsc{scalar} SvS&     32.0   &        2.8    &        32.0   &        3.3 & 32.0 & 6.6 \\

\end{tabular}
\end{table}

\subsection{Bitmap-list hybrids (\textsc{hyb+m2})}
\label{sec:exphyb}

We can also use bitmaps to accelerate intersections.
Culpepper and Moffat's \textsc{hyb+m2} framework is simple but effective~\cite{Culpepper:2010:ESI:1877766.1877767}: all lists where the average gap size is smaller or equal to $B$ (where $B=8,16\text{ or }32$) are stored as
bitmaps whereas other lists are stored as compressed deltas (using \textsc{varint} or
another compression algorithm).  To compute any intersection, the compressed lists are first
intersected (e.g., using galloping) and the result is then checked against
the bitmaps one by one.
It is a hybrid approach in the sense that it uses both bitmaps and compressed deltas.
We  modify their framework by replacing the compression and intersection functions with SIMD-based ones. To simplify our analysis, we use
galloping intersections with scalar compression schemes such as \textsc{varint} and \textsc{FastPFOR}, and SIMD intersections  with SIMD compression schemes.
Intersections between lists are always computed over uncompressed lists. SIMD intersections are computed using the hybrid made of \textsc{V1}, \textsc{V3} and \textsc{SIMD Galloping} presented in \S~\ref{sec:hybrid}.

To improve data cache utilization, we split our corpora into parts processed separately (32~parts for GOV2 and 64~parts for ClueWeb09).
The partition is based on document identifiers: we have, effectively, 32 or 64~independent indexes that apply to different document subsets.
Thus, all intermediate results for a single part can fit into L3 cache.
Given a query, we first obtain the result for the first part, then for the second, and so on.
Finally, all partial results are collected in one array.
In each part, we  work directly on compressed lists and bitmaps, without other auxiliary data structures. The lists are decompressed and then intersected. We expect our approach to be reasonably efficient given our benchmark: in Appendix~\ref{appendix:hybskip}, we validate it against an alternative that involves skipping.

\begin{figure}
\centering
\subfloat[\textsc{GOV2} (unsorted)]{\centering\includegraphics[width=0.7\columnwidth]{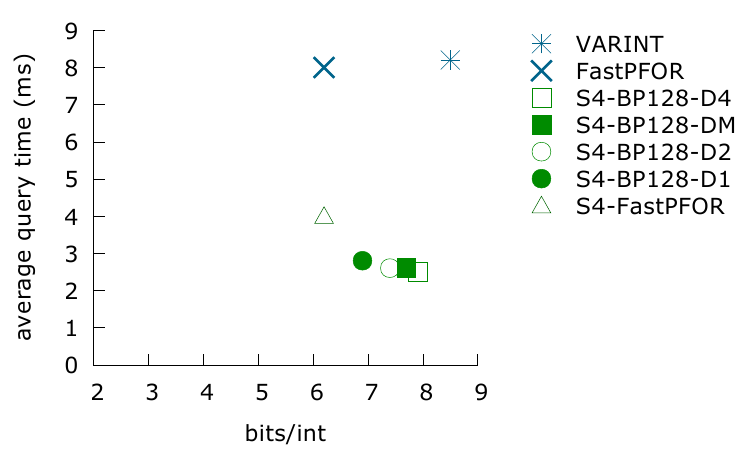}}\\
\subfloat[\textsc{GOV2} (sorted)]{\centering\includegraphics[width=0.7\columnwidth]{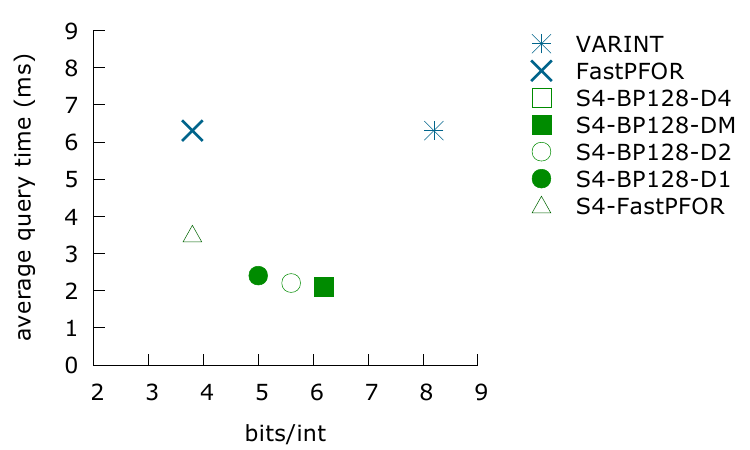}}\\
\subfloat[\textsc{ClueWeb09}]{\centering\includegraphics[width=0.7\columnwidth]{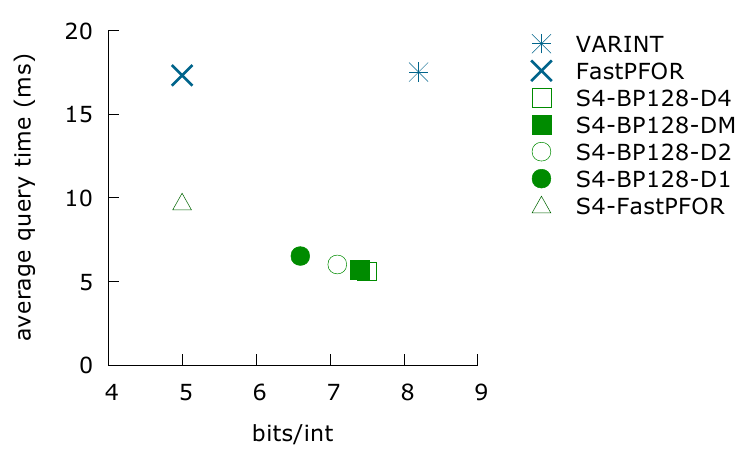}}
\caption{\label{fig:nobitmap}Average time required to answer queries  vs.\ the storage requirement in bits/int %using partitions but no bitmap.
without bitmap
}
\end{figure}

As shown in  Table~\ref{table:hybresults} and Fig.~\ref{fig:nobitmap},
the schemes using SIMD instructions are 2--3 times
faster than \textsc{FastPFOR} and \textsc{varint} while having comparable
compression ratios.
The retrieval times can be improved (up to  $\approx 4\times$)
with the addition of bitmaps, but  the index is sometimes larger (up to  $\approx 2\times$).

%The partitioning strategy always improved performance of non-hybrid methods by 10--\SI{40}{\percent}
%without affecting compression rates.
%Thus, in this case, we include only results for partitioned indices.
%However, for hybrid schemes, partitioning can improve performance, compression
%ratios, or both characteristics.
%Thus, we present two set of results: for partitioned
%and non-partitioned indices.
%Performance of partitioned and non-partitioned bit\-maps is different,
%because each part of the posting list may be encoded differently.
%For example, even though the average gap size
%for the complete posting is smaller or equal than $B$,
%after partitioning, there may be parts with the average gap larger than $B$.
%Unlike the unpartitioned posting, which is stored as a bitmap,
%such parts are stored as compressed deltas, thus using less space.

%Ok
%\daniel{I have removed the non-partitioned case. It is generally slower.}
For $B=8$, $B=16$, and indices without bitmaps, the
various \textsc{S4-BP128-*} schemes provide different space vs.~performance trade offs.
For example, for ClueWeb09 and $B=8$%(partitioned)
, \textsc{S4-BP128-D4}
is \SI{11}{\percent} faster than \textsc{S4-BP128-D1}, but \textsc{S4-BP128-D1}  uses \SI{7}{\percent} less space.

\begin{figure}\centering
\includegraphics[width=0.7\columnwidth]{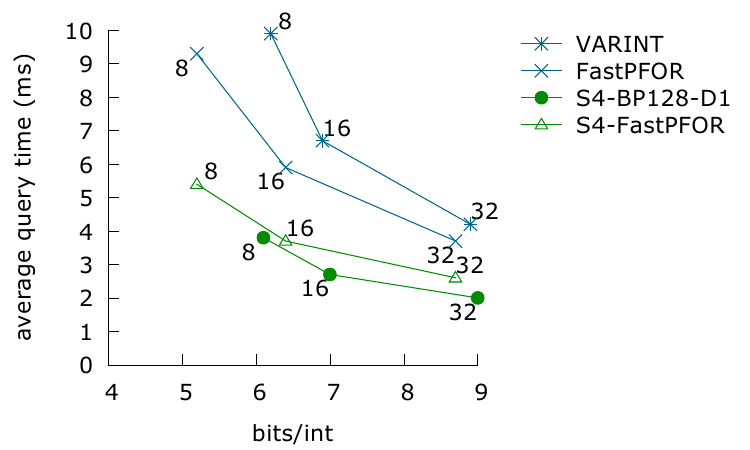}
\caption{\label{fig:cluewebbitmaps}Average time required to answer queries  vs.\ the storage requirement in bits/int using partitions and various bitmap parameters ($B=8,16,32$) for  \textsc{ClueWeb09}.}
\end{figure}

We also compare scalar vs.\ SIMD-based algorithms. For ClueWeb09, \textsc{S4-FastPFOR-D1}  can be nearly twice as fast as the non-SIMD \textsc{FastPFOR} scheme (see Fig.~\ref{fig:cluewebbitmaps}). Even with \mbox{$B=16$} where we rely more on the bitmaps for speed, \textsc{S4-FastPFOR} is over \SI{50}{\percent} faster than \textsc{FastPFOR} (for ClueWeb09).
As expected, for $B=32$, there is virtually no difference in speed and compression ratios among various \textsc{S4-BP128-*} compression schemes.
However,
\textsc{S4-BP128-D4}  is still about twice as fast as \textsc{varint} so that, even in this extreme case, our SIMD algorithms are worthwhile.

As a reference point, we also implemented intersections using the Skipper~\cite{sanders2007intersection} approach. In this last case, posting lists are stored as deltas compressed using \textsc{varint}. However, to support random access, we have an auxiliary data structure containing sampled uncompressed values with pointers inside the array of compressed deltas. We found that sampling every 32~integers gave good results. Because we need to store both the uncompressed integer and a pointer for each 32~integers, the auxiliary Skipper data structure adds about two bits of storage per integer. We refer to the original paper and to our software for details. We find
that Skipper was 2 to 3~times faster than intersections over \textsc{varint}-compressed deltas (at the expense of storage) without bitmap, but compared with the \textsc{hyb+m2} framework ($B=8,16,32$) or our SIMD-based approaches, it is not competitive. Though we could replace \textsc{varint} in Skipper by a faster alternative, we expect that the benefits of skipping would then be diminished.

We also recorded the median and the $90^{\textrm{th}}$~percentile
retrieval times. All these metrics benefited from our techniques. Moreover, our measures are accurate: after repeating experiments five times, we find that all standard deviations are less than \SI{1}{\percent}.

\begin{table}
\caption{\label{table:hybresults}Time required to answer queries from TREC~1MQ in  ms/query along with the storage requirement in bits/int. Entropies of the deltas are provided. We write
\textsc{S4-FastPFOR} as a shorthand for
\textsc{S4-FastPFOR-D1}. We omit \textsc{S4-BP128-DM} when it does not differ significantly from \textsc{S4-BP128-D4}.
 }
\centering

\begin{tabular}{l@{\hspace{-0.2em}}c@{\hspace{0.5em}}c@{\hspace{0.5em}}c@{\hspace{0.5em}}c@{\hspace{0.5em}}c@{\hspace{0.5em}}c@{\hspace{0.1em}}}\hline
scheme &
bits/int & time &
bits/int & time &
bits/int & time  \\\hline\hline
&
\multicolumn{2}{@{\hspace{0em}}c@{\hspace{0em}}}{\textsc{GOV2} } &
\multicolumn{2}{@{\hspace{0em}}c@{\hspace{0em}}}{\textsc{GOV2} } &
\multicolumn{2}{@{\hspace{0em}}c@{\hspace{0em}}}{\textsc{ClueWeb09} } \\
&
\multicolumn{2}{@{\hspace{0em}}c@{\hspace{0em}}}{(sorted) } &
\multicolumn{2}{@{\hspace{0em}}c@{\hspace{0em}}}{(unsorted) } &
\multicolumn{2}{@{\hspace{0em}}c@{\hspace{0em}}}{} \\
\hline

entropy & \multicolumn{2}{c} {1.9} & \multicolumn{2}{c}{4.6} & \multicolumn{2}{c}{3.1}\\

\hdashline[1pt/1pt]\multicolumn{7}{c}{\textsc{no bitmap}%(partitioned)
}\\
\hdashline[1pt/1pt]
\textsc{varint}&         8.2    &        6.3    &        8.5    &        8.2 & 8.2 & 17.5 \\
\textsc{FastPFOR}&       3.8    &        6.3    &        6.2    &        8.0 & 5.0 & 17.3 \\
\textsc{S4-BP128-D4}&     6.2    &        2.1    &        7.9    &        2.5 & 7.5 & 5.6 \\
\textsc{S4-BP128-DM}&     6.2    &        2.1    &        7.7    &        2.6 & 7.4 & 5.7 \\
\textsc{S4-BP128-D2}&     5.6    &        2.2    &        7.4    &        2.6 & 7.1 & 6.0 \\
\textsc{S4-BP128-D1}&     5.0    &        2.4    &        6.9    &        2.8 & 6.6 & 6.5 \\
\textsc{S4-FastPFOR}&    3.8    &        3.5    &        6.2    &        4.0 & 5.0 & 9.7 \\
\\
\hdashline[1pt/1pt]\multicolumn{7}{c}{$B=8$
 %(partitioned)
 } \\\hdashline[1pt/1pt]
\textsc{varint}&         5.8    &        3.0    &        7.4    &        4.7 & 6.2 & 9.9 \\
\textsc{FastPFOR}&       4.3    &        3.0    &        6.3    &        4.5 & 5.2 & 9.3 \\
\textsc{S4-BP128-D4}&     5.7    &        1.2    &        7.4    &        1.6 & 6.5 & 3.4 \\
\textsc{S4-BP128-DM}&     5.7    &        1.2    &        7.3    &        1.6 & 6.4 & 3.4 \\
\textsc{S4-BP128-D2}&     5.5    &        1.3    &        7.1    &        1.6 & 6.3 & 3.6 \\
\textsc{S4-BP128-D1}&     5.3    &        1.4    &        6.8    &        1.7 & 6.1 & 3.8 \\
\textsc{S4-FastPFOR}&    4.3    &        1.9    &        6.3    &        2.3 & 5.2 & 5.4 \\
\hdashline[1pt/1pt]\multicolumn{7}{c}{$B=16$
%(partitioned)
} \\\hdashline[1pt/1pt]
\textsc{varint}&         6.3    &        2.1    &        8.2    &        3.2 & 6.9 & 6.7 \\
\textsc{FastPFOR}&       5.5    &        2.1    &        7.6    &        2.9 & 6.4 & 5.9 \\
\textsc{S4-BP128-D4}&     6.5    &        0.9    &        8.4    &        1.2 & 7.2 & 2.5 \\
\textsc{S4-BP128-D2}&     6.4    &        1.0    &        8.1    &        1.2 & 7.1 & 2.6 \\
\textsc{S4-BP128-D1}&     6.2    &        1.1    &        7.9    &        1.3 & 7.0 & 2.7 \\
\textsc{S4-FastPFOR}&    5.5    &        1.4    &        7.6    &        1.6 & 6.4 & 3.7 \\
\hdashline[1pt/1pt]\multicolumn{7}{c}{$B=32$
%(partitioned)
} \\\hdashline[1pt/1pt]
\textsc{varint}&         8.2    &        1.4    &        11.1   &        2.0 & 8.9 & 4.2 \\
\textsc{FastPFOR}&       7.8    &        1.5    &        10.9   &        1.8 & 8.7 & 3.7 \\
\textsc{S4-BP128-D4}&     8.4    &        0.8    &        11.2   &        0.9 & 9.1 & 1.9 \\
\textsc{S4-BP128-D2}&     8.4    &        0.8    &        11.1   &        1.0 & 9.1 & 2.0 \\
\textsc{S4-BP128-D1}&     8.3    &        0.8    &        11.0   &        1.0 & 9.0 & 2.0 \\
\textsc{S4-FastPFOR}&    7.8    &        1.1    &        10.9   &        1.2 & 8.7 & 2.6 \\
\\
\hdashline[1pt/1pt]
Skipper~\cite{sanders2007intersection} &         10.2   &        2.6    &        10.5   &        4.3 & 10.2 & 7.7 \\ \hdashline[1pt/1pt]

\end{tabular}
\end{table}

\section{Discussion and conclusions}

We demonstrated that combining unpacking and differential coding
resulted in faster decompression speeds,
which were approximately \SI{30}{\percent} better
than the best speeds reported previously~\cite{LemireBoytsov2013decoding}.
%For example, \textsc{S4-BP128-D1} can decode integers at about  1.2~integers per CPU cycle. To promote its use among practitioners and researchers, we have published a corresponding C library\footnote{\protect\url{https://github.com/lemire/simdcomp}} under a liberal license.
To match the performance of these fast compression schemes,
we additionally vectorized and optimized the intersection of posting lists.
To this end, we introduced a  family of algorithms exploiting commonly available SIMD instructions (\textsc{V1}, \textsc{V3} and \textsc{SIMD Galloping}). They are often twice as fast as the best non-SIMD algorithms.
Then, we used our fast SIMD routines for decompression and posting intersection
 to accelerate Culpepper and Moffat's \textsc{hyb+m2}~\cite{Culpepper:2010:ESI:1877766.1877767}.
We believe that \textsc{hyb+m2} is one of the fastest published algorithms  for conjunctive queries.
Yet we were able to sometimes double the speed of \textsc{hyb+m2} without sacrificing compression.

%Our software only uses SSE2 instructions dating back to the Pentium~4. Because we do
%not use specialized instructions,
%we expect that some of our good results are portable to other platforms such as ARM NEON and IBM~POWER8.
%% This warrants further investigation.
% \daniel{Of course, you can say that we use old instructions as if it were a bad thing. But it can also be a good thing.}
% However,
%\textsc{hyb+m2} has limitations:
%e.g,  it is
%unclear how it could extended to support scoring efficiently (e.g., using BM25).
%In future work, we will apply our fast SIMD compression and intersection techniques
%together with multicore processing~\cite{Tatikonda:2011:PLI:2009916.2010045} and  scoring functions~\cite{Catena2014}. We are also interested in how our techniques scale to billions of documents: in such
%cases, we would need  64-bit document identifiers.

%We have shown that  vectorized compression and intersection could be so fast that---without any auxiliary data structure---it is possible to surpass the speed of a fast scalar intersection scheme (galloping) by a wide margin. We believe that only by using auxiliary data structures were researchers able to achieve similar results. However, it should be possible to get even better results by using SIMD-optimized data structures as in Schlegel et al.~\cite{Schlegel2011}.

Our work was focused on 128-bit vectors.
Intel and AMD recently released processors that support integer
operations on 256-bit vectors using the new AVX2 instruction set.
On such a processor, Willhalm et al.~\cite{WillhalmO0F13} were able to double their bit unpacking speed.
Moreover, Intel plans to support 512-bit vectors in 2015 on its commodity processors. Thus
optimizing algorithms for SIMD instructions will become even more
important in the near future.

%%%%%%%%%%%%%%%%%%%%%
% Silenced for submission as per Leo's request
%%%%%%%%%%%%%%%%%%%%%%
\begin{comment}
\begin{anonymous}\section{Acknowledgments}
D.~Lemire acknowledges support from the Natural Sciences and Engineering Research
Council of Canada (NSERC) from grant number 26143. We thank T.~Wu from Genentech for his comments on an earlier version of the manuscript. We thank O.~Kaser from UNB for code contributions. We thank J.~Callan from CMU for his help.\end{anonymous}
\end{comment}

\bibliographystyle{wileyj}
\bibliography{bib/acmtitles.bib,bib/lemur.bib}
\appendix

\section{Sample C++ code using Intel intrinsics for intersection algorithms}\label{appendix:code}

\lstset{language=C, showstringspaces=false, breaklines,
 emph={p}, emphstyle={\textbf},
 emph={[2]endp}, emphstyle={[2]\textbf},
emph={[3]m}, emphstyle={[3]\textbf},
emph={[4]sum}, emphstyle={[4]\textbf},
emph={[5]uint32}, emphstyle={[5]\textbf},
emph={[6]uint64}, emphstyle={[6]\textbf},
emph={[7]__m128i}, emphstyle={[7]\textbf}
}

\begin{lstlisting}
uint32_t *r // pointer to short list
uint32_t *f // pointer to long list

// load T = 8 integers from f
__m128i F = _mm_loadu_si128((__m128i*)f);
__m128i G = _mm_loadu_si128((__m128i*)f+1);

// replicate current value from r
__m128i R = _mm_set1_epi32(*r);

// compare R and F, G
__m128i T0 =  _mm_cmpeq_epi32(F,R);
__m128i T1 =  _mm_cmpeq_epi32(G,R);
T = _mm_or_si128 (T0,T1);
/*
  if SSE4 is supported, we can instead use
  a slightly faster instruction _mm_testz_si128(T, T)
*/
if (_mm_movemask_epi8(T) != 0) {
  //one value in R matches a value in F, G
}
\end{lstlisting}

\section{Bitmap-list hybrids (\textsc{hyb+m2}) with skipping}\label{appendix:hybskip}

In \S~\ref{sec:exphyb}, we describe experiments using the \textsc{hyb+m2} model.
Short lists are first decompressed and then intersected (e.g., using the galloping algorithm).
As recommended by Culpepper and Moffat~\cite{Culpepper:2010:ESI:1877766.1877767},
we do not use skipping~\cite{moffat1996self}: all integers from the short lists are decompressed.
In contrast,
Kane and Tompa~\cite{kane2014} found skipping over large blocks in the \textsc{hyb+m2} context to be highly beneficial.
For \textsc{varint} only, we adopt the Kane-Tompa approach, replacing galloping and partitions by skipping with blocks of 32 and 256~integers, and
present the results in Table~\ref{table:varintskiphybresults} using the same tests as in \S~\ref{sec:exphyb}.
We see that skipping is worse than galloping in most cases (up to \SI{60}{\percent}) though it can be sometimes be slightly beneficial (up to \SI{15}{\percent}).
Though these results are only for \textsc{varint} compression, we expect the benefits of skipping to be less for faster compression schemes such as \textsc{S4-BP128-D4}.

\begin{table}
\caption{\label{table:varintskiphybresults} Time required to answer queries  from TREC~1MQ using bitmaps (\textsc{hyb+m2}) and short lists compressed with  \textsc{varint}.
We compare a skipping approach over short lists using blocks of 32 or 256 integers (as in Kane and Tompa~\cite{kane2014}) vs.\ an approach using galloping over partitioned data.
 }
\centering

\begin{tabular}{lccc}\hline
 &
GOV2 & GOV2 & \textsc{ClueWeb09} \\
 &
sorted & unsorted &  \\\hline
\hdashline[1pt/1pt]\multicolumn{4}{c}{$B=8$
 %(partitioned)
 } \\\hdashline[1pt/1pt]
galloping &            3.0    &  4.7 & 9.9 \\
skipping (32) &        4.3    &  7.3 & 11\\
skipping (256) &       4.4    & 7.5  & 12\\

\hdashline[1pt/1pt]\multicolumn{4}{c}{$B=16$
%(partitioned)
} \\\hdashline[1pt/1pt]
galloping &              2.1    & 3.2 & 6.7\\
skipping (32) &          2.5    & 4.2 & 6.4\\
skipping (256) &         2.6    & 4.3 & 6.9\\

\hdashline[1pt/1pt]\multicolumn{4}{c}{$B=32$
%(partitioned)
} \\\hdashline[1pt/1pt]
galloping&         1.4    & 2.0 & 4.2 \\
skipping (32) &    1.3    & 2.1 & 3.6\\
skipping (256) &   1.4    & 2.1 & 3.8\\

\end{tabular}
\end{table}

\label{theend}
\end{document}